\documentclass[10pt,preprint]{aastex}

\newcommand{\dsf}[2]{\displaystyle{\frac{#1}{#2}}}
\slugcomment{\scriptsize{ACCEPTED TO ApJ, Preprint typset using \LaTeX style}}
\shorttitle{MHD INSTABILITY IN PROTO-NEUTRON STARS (II)}
\shortauthors{Masada et al.}
\begin{document}
\title{The Effect of Neutrino Radiation on Magnetorotational Instability in 
Proto-Neutron Stars}

\author{Youhei Masada\altaffilmark{1,2}, Takayoshi Sano\altaffilmark{3},
  and Kazunari Shibata\altaffilmark{1}}

\altaffiltext{1}{Kwasan and Hida Observatories, Kyoto University, Yamashina, 
Kyoto 607-8471, Japan; masada@kusastro.kyoto-u.ac.jp}
\altaffiltext{2}{Department of Astronomy, Kyoto University, Sakyo, Kyoto 606-8502, Japan}
\altaffiltext{3}{Institute of Laser Engineering, Osaka University, Suita, 
Osaka 565-0871, Japan} 
\begin{abstract}
Neutrino radiation takes a major role in the momentum, heat, and lepton 
transports in proto-neutron stars (PNSs). 
These diffusive processes affect the growth of 
magnetorotational instability (MRI) in PNSs. 
We perform a local linear analysis for the axisymmetric and 
nonaxisymmetric MRI including the effects of 
neutrino transports and ohmic dissipation. 
We find that the MRI can grow even in the multi-diffusive 
situations that are realized in neutrino loaded PNSs. 
When the toroidal magnetic component dominates over 
the poloidal one, nonaxisymmetric MRI modes grow much faster than 
axisymmetric modes. These results suggest the importance of the nonaxisymmetric 
MRI in PNSs. Thus the understandings of three-dimensional nonlinear evolutions 
of the MRI are necessary to reveal the explosion mechanism of core-collapse supernovae. 
\end{abstract}

\section{INTRODUCTION}
Strong magnetic fields can play an important role on the dynamics 
of core-collapse supernovae. Akiyama et al. (2003) pointed out the 
possibility of the field amplification by the magnetorotational instability (MRI) 
in supernova core. The shapes of shock wave and the neutrinosphere are 
modified aspherically by the coupling effects of the rotation and magnetic fields, 
and anisotropic neutrino radiations could lead to jet-like explosions (Kotake et al.2003, 2004). 
Magnetic pressure driven explosions are also investigated in the context of 
magneto-rotational core collapse (Yamada \& Sawai 2004; Mizuno et al. 2004; Takiwaki et al. 2005; 
Ardeljan et al. 2005; Sawai et al. 2005; Moissenko et al. 2006). 
In these studies, strong magnetic fields ($ \gtrsim 10^{15} \ \rm{G}$) 
are generated by the field compression, field wrapping, and the MRI, 
which lead to the prompt explosions of core collapse supernovae. 
Alternative effects of magnetic fields are proposed by Thompson et al.(2005). 
They suggest that the turbulent viscosity sustained by the MRI converts 
the rotational energies of supernova core into the thermal energies, 
and thus the MRI will affect the dynamics of core collapse supernovae. 
However, these studies mainly focus on the dynamical effects of magnetic fields 
at the outside of PNSs. Therefore, magnetic phenomena inside PNSs and 
its influence on the core collapse dynamics have not been understood yet. 

There are a few studies of magnetic effects at the interiors of PNSs. 
Miralles et al. (2002) show that the convective fluid motions in PNSs 
are prevented by strong magnetic fieds. 
Numerical simulations suggest that the magnetic bubbles rise up by buoyancy 
and carry neutrino-rich materials to the neutron star surface (Wilson et al. 2005). 
This could increase the neutrino luminosity sufficietly 
to achieve a successful neutrino driven explosion. 
Socrates et al.(2005) show that the neutrino bubble instability 
grows at the inside of magnetized PNSs. 
Nonlinear evolution of the neutrino bubble instability will 
generate large density fluctuations. The radiation preferentially 
leaks out of the rarefied regions, which enhances the outgoing flux. 
They also propose that these global asymmetry 
in the neutino emission would cause the pulsar kicks. 

Masada et al. (2006; hearafter Paper I) investigate 
the growth of the nonaxisymmetric MRI 
and show that it can grow even in the stably stratified envelope of PNSs. 
Our results suggest that the nonlinear evolution of MRI amplifies the magnetic fields 
and drives MHD turbulence in PNSs. The layers below the neutrinosphere are thought 
to be convectively stable. If buoyant motions are excited by the MRI,  
it could lead to enhancement of the neutrino luminosity. 
However, in Paper I, the effects of neutrino radiations on the growth of MRI are 
neglected. Below the neutrinosphere, electron-type neutrinos exchange energy 
with the matter via pair capture process and exchange momentum via 
elastic neutirno-nucleon scattering. 
These processes dominate the other 
transport processes by electrons and/or photons in PNSs, 
and thus the coefficients of the momentum, heat, and lepton transports are 
determined by neutrino radiative processes. 
The importance of these transports caused by 
the neutrino radiation have been suggested already in the previous studies of 
the mixing processes in PNSs (Bruenn \& Dinneva 1996; Mezzacappa et al. 1998; 
Miralles et al. 2000; Bruenn et al. 2004; Socrates et al. 2005). 
Therefore, we must study such neutrino effects to reveal 
the growth of MRI in more realistic, neutrino loaded PNSs. 

In this paper, we analyze the diffusive effects caused by 
neutrino radiations on the growth of MRI and 
discuss the natures of MRI in PNSs. 
We consider four kinds of diffusive processes; the momentum, heat and chemical 
diffusions induced by the neutrino radiation and the ohmic dissipation of hot nuclear matters. 
As the mean free path of neutrino is short compared to the 
density and temperature length scales below the neutrinosphere, 
we can use diffusion approximation for the neutrino transports (Socrates et al. 2005). 
Another important feature of our analysis is to consider 
the vertical magnetic field as well as the toroidal one. 
For this case, both axisymmetric and nonaxisymmetric modes 
can be unstable for the MRI. In previous works, axisymmetric and 
nonaxisymmetric MRI are investigated separately 
(Acheson 1978; Balbus \& Hawley 1994; Menou et al. 2004). 
We consider both axisymmetric and nonaxisymmetric stability 
and discuss their difference in the growth rate and stability conditions. 

The paper is organized as follows. 
In \S~2 we obtain the dispersion equation including diffusive processes, 
which determines the stability of magnetized PNSs. 
Necessary conditions for the stability are derived by an analytical approach. 
In \S~3 we discuss general properties of the dispersion relation 
focusing on the effects of quadruple diffusivities. 
In \S~4 we apply our results to the interior of magnetized PNSs, 
and examine the effects of neutrino radiations on the growth of MRI. 
We also examine the difference between axisymmetric and 
nonaxisymmetric modes of MRI. The possibility of the other instabilities 
is discussed in \S~5. Finally, we summarize our main findings in \S~6.

\section{LOCAL DISPERSION EQUATION}
\subsection{Physical assumptions and basic equations}
We examine the stability of differentially rotating, magnetized PNSs 
using a linear perturbation theory. 
In this paper, we focus on the effects of the neutrino radiation 
for MHD instabilities. The weak interaction between nuclear 
matters and neutrinos is described by using various opacities. 
We simply assume that the energy and momentum are exchanged between 
electron-type neutrinos and nuclear matters via pair capture process 
and elastic neutrino-nucleon scattering. In addition, 
neutrino opacities are assumed to be proportional to the square of 
the neutrino energy $E$, 
\begin{equation}
\kappa_{\nu} = \kappa_0 (E/E_0)^2 \;, \label{eq1}
\end{equation}
where $\kappa_0$ and $E_0$ are constants and 
independent of composition (e.g., Socrates et al.2005). 
In this paper, opacities are defined as the inverse of mean free paths. 

Neutrinos are optically thick and both the thermal and chemical equilibrium 
would be held below the neutrinosphere. Thus, the governing equations are the following,
\begin{equation}
\displaystyle{\frac{\partial\rho}{\partial t}} 
 + \nabla \cdot (\rho \mbox{\boldmath $u$}) = 0 \;, 
\label{eq2}
\end{equation}
\begin{equation}
\displaystyle{\frac{\partial \mbox{\boldmath $u$}}{\partial t}} 
+ ( \mbox{\boldmath $u$}\cdot
\nabla ) \mbox{\boldmath $u$} = - \frac{1}{\rho }\nabla
\left( P +\frac{1}{8\pi }\mbox{\boldmath $B$}^2 \right) 
+ \frac{1}{4\pi\rho}( \mbox{\boldmath $B$}\cdot\nabla ) 
\mbox{\boldmath $B$} + \mbox{\boldmath $g$} + \nu\nabla^2\mbox{\boldmath $u$}\;,
\label{eq3} 
\end{equation}
\begin{equation}
\displaystyle{\frac{\partial\mbox{\boldmath $B$}}{\partial t}} = 
\nabla \times (\mbox{\boldmath $u$}\times\mbox{\boldmath $B$}) 
+ \eta\nabla^2\mbox{\boldmath $B$}\;, 
\label{eq4} 
\end{equation}
\begin{equation}
n\Big( \displaystyle{\frac{\partial Y_L}{\partial t}} 
+ \mbox{\boldmath $u$} \cdot \nabla Y_L \Big) = 
- \nabla\cdot\mbox{\boldmath $F$}_L \;, \label{eq5}
\end{equation}
\begin{equation}
nT\Big( \displaystyle{\frac{\partial s}{\partial t}} 
+ \mbox{\boldmath $u$} \cdot \nabla s \Big) + 
n\mu_{\nu_e} \Big( \displaystyle{\frac{\partial Y_L}{\partial t}} 
+ \mbox{\boldmath $u$} \cdot \nabla Y_L \Big) =  
- \nabla\cdot\mbox{\boldmath $F$} \;. \label{eq6}
\end{equation}
These are the continuity equation, equation of motion, 
induction equation, conservations of lepton fraction and energy, respectively. 
Here $\rho$ is the fluid density, $\mbox{\boldmath $u$}$ is the flow velocity, 
$P$ is the total pressure of nuclear matters and radiations, 
$\mbox{\boldmath $B$}$ is the magnetic field, 
$\mbox{\boldmath $g$}$ is the gravitational field, 
$n$ is the baryon number density, $s$ is the entropy per baryon, 
$Y_L$ is the lepton fraction, $\mu_{\nu_e}$ is the neutrino chemical potential, 
and $\mbox{\boldmath $F$}$ and $\mbox{\boldmath $F$}_L $ are the radiative energy flux 
and lepton flux, respectively. 
Neutrino viscosity and magnetic diffusivity are 
represented by $\nu$ and $\eta$. The effect of the bulk viscosity is neglected here. 

We employ the equilibrium diffusion approximation, and thus 
our analysis is restricted to the region below the neutrinosphere. 
Using the energy flux of the electron neutrino and anti-neutrino, 
$\mbox{\boldmath $F$}_{\nu_e} $ and $\mbox{\boldmath $F$}_{\bar{\nu_e}}$, 
the radiative energy flux $\mbox{\boldmath $F$}$ is given by 
\begin{equation}
\mbox{\boldmath $F$} = 
\mbox{\boldmath $F$}_{\nu_e} + \mbox{\boldmath $F$}_{\bar{\nu_e}} 
=  - a_T \nabla T - a_L \nabla Y_L \label{eq7} \;,
\end{equation}
where 
\begin{equation}
a_T = \displaystyle{\frac{E_0^2}{6\kappa_0 \hbar^3 c^2}}\displaystyle{\frac{k_B^2 T}{3}} \;, \ \ \ 
a_L = \displaystyle{\frac{E_0^2}{6\kappa_0 \hbar^3 c^2}}\displaystyle{\frac{\mu_{\nu_e}}{\pi^2}}
\Big(\displaystyle{\frac{\partial \mu_{\nu_e}}{\partial Y_L}} \Big)_{P,T} \label{eq8} \;
\end{equation} 
(Bludman \& Van Riper 1978). The lepton flux is given by 
\begin{equation}
\mbox{\boldmath $F$}_L = -  b_T\nabla T - b_L \nabla Y_L  \;, \label{eq9}
\end{equation}
where  
\begin{equation}
b_L = a_L/\mu_{\nu_e} = \displaystyle{\frac{E_0^2}{6\kappa_0 \hbar^3 c^2}}
\displaystyle{\frac{1}{\pi^2}}
\Big(\displaystyle{\frac{\partial \mu_{\nu_e}}{\partial Y_L}} \Big)_{P,T} \label{eq10} \;. 
\end{equation}
Using the simple parameterization of opacity defined by equation~(\ref{eq1}), 
the temperature gradient cannot contribute to the lepton flux, 
that is $b_T = 0$ (Socrates et al. 2005). 
Using equations (\ref{eq7}) and (\ref{eq9}), we can rewrite the 
conservations of lepton fraction and energy; 
\begin{equation}
n\Big( \displaystyle{\frac{\partial Y_L}{\partial t}} 
+ \mbox{\boldmath $u$} \cdot \nabla Y_L \Big) = 
b_L \nabla^2 Y_L \;, \label{eq11}
\end{equation}
\begin{equation}
nT\Big( \displaystyle{\frac{\partial s}{\partial t}}  
+ \mbox{\boldmath $u$} \cdot \nabla s \Big) = a_T \nabla^2 T \;. \label{eq12}
\end{equation}
We ignore the spatial dependence of the diffusion coefficients $a_T$, $a_L$, 
and $b_L$. This is appropriate for a local linear analysis. 
The resistive and viscous dissipation terms are also neglected in the energy 
equation, because they are higher order terms. 

\subsection{Linearized Equations}
Following the analysis performed in Paper I, we consider
the Eulerian perturbations (denoted 
by a prefix $\delta $) with the WKB spatial and temporal dependence, 
$\delta \propto \exp \{i (k_r r + m\phi + k_z z - \sigma t)\}$. 
Local WKB analysis is irrelevant in the cases that the growth rate of 
nonaxisymmetric unstable modes is smaller than the differential rotation rate. 
However, we now focus mainly on the MRI in the cases with strong velocity shear 
for which the growth rate is comparable to the angular velocity 
and adopt the simple WKB analysis in this paper. 
We use the cylindrical coordinates $(r, \phi, z)$ and consider a PNS rotating 
with the angular velocity $\Omega(r)$, and 
its magnetic fields being $\mbox{\boldmath $B$} = (0,B_{\phi},B_z )$. 
We assume magnetic fields are uniform locally for simplicity. 
At the unperturbed state, the PNS is in the magnetohydrostatic 
equilibrium, that is, 
\begin{equation}
\displaystyle{\frac{B_\phi^2}{4\pi\rho r}} + \displaystyle{\frac{1}{\rho}}\displaystyle{\frac{\partial P}{\partial r}} = 
-g_r + r\Omega^2 \;,
\ \ \ \ \ \displaystyle{\frac{1}{\rho}}
\displaystyle{\frac{\partial P}{\partial z}} = -g_z \;,
\label{eq13}
\end{equation}
in the radial and vertical directions, respectively. 
Written out in component form and the largest terms retained, 
equations (\ref{eq2})-(\ref{eq4}) become to linear order: 
\begin{equation}
k_r\delta u_r + k_z\delta u_z  =  0 \;, 
\label{eq14} 
\end{equation}
\begin{equation}
i\omega_{\nu}\delta u_r + 2\Omega\delta u_{\phi} +
\displaystyle{\frac{i(\mbox{\boldmath $k$}\cdot \mbox{\boldmath $B$})}{4\pi\rho}}\delta b_r = 
\displaystyle{\frac{ik_r}{\rho}}\delta P + 
\displaystyle{\frac{ik_r(\mbox{\boldmath $B$}\cdot
\delta\mbox{\boldmath $b$})}{4\pi\rho}} + \displaystyle{\frac{B_{\phi}\delta b_{\phi}}{2\pi\rho r}}
- \displaystyle{\frac{\delta\rho}{\rho}}(g_r -r\Omega^2) \;, 
\label{eq15} 
\end{equation}
\begin{equation}
i\omega_{\nu}\delta u_{\phi} - \displaystyle{\frac{1}{r}}\displaystyle{\frac{\partial (r^2\Omega)}{\partial r}}\delta u_r 
+ \displaystyle{\frac{i(\mbox{\boldmath $k$}\cdot\mbox{\boldmath $B$})}{4\pi\rho}}\delta b_{\phi} 
+ \displaystyle{\frac{B_{\phi}\delta b_r}{4\pi\rho r}} = 0 \;, 
\label{eq16} 
\end{equation}
\begin{equation}
i\omega_{\nu}\delta u_z + \displaystyle{\frac{i(\mbox{\boldmath $k$}\cdot 
\mbox{\boldmath $B$})}{4\pi\rho}}\delta b_z 
= \displaystyle{\frac{ik_z}{\rho}}\delta P + 
\displaystyle{\frac{ik_z(\mbox{\boldmath $B$}\cdot\delta\mbox{\boldmath $b$})}{4\pi\rho}} -
\displaystyle{\frac{\delta\rho}{\rho}}g_z \;,
\label{eq17} 
\end{equation}
\begin{equation}
\omega_{\eta}\delta b_r + (\mbox{\boldmath $k$}\cdot\mbox{\boldmath $B$})\delta u_r = 0\;,
\label{eq18} 
\end{equation}
\begin{equation}
i\omega_{\eta}\delta b_{\phi} + i(\mbox{\boldmath $k$}\cdot\mbox{\boldmath $B$})\delta u_{\phi} 
+ r \displaystyle{\frac{\partial\Omega}{\partial r}}\delta b_r + \displaystyle{\frac{B_{\phi}}{r}}\delta u_r = 0\;, 
\label{eq19} 
\end{equation}
\begin{equation}
\omega_{\eta}\delta b_z + (\mbox{\boldmath $k$}\cdot\mbox{\boldmath $B$}) \delta u_z = 0\;,
\label{eq20}
\end{equation}
where  
$$
\omega = \sigma - m\Omega \;, \ \ 
\omega_{\nu} = \omega + i\nu k^2 \;, \ \ \omega_{\eta} = \omega + i\eta k^2 \;,
$$
$$
k^2 = k_r^2 + k_z^2 \;,\ \ 
\mbox{\boldmath $k$} = (k_r, m/r, k_z) \;, \ \ 
\delta\mbox{\boldmath $b$} = (\delta b_r,\delta b_\phi , \delta b_z)\;.
$$
Here we use the local approximation $(k \gg 1/r, m/r)$. 
We also adopt the Boussinesq approximation 
because the typical speeds considered in our analysis are 
much smaller than the sound speed. 
The relation $\delta P + B_{\phi}\delta b_{\phi}/4\pi  = 0$ 
is used in equation~(\ref{eq16}) by virture of 
the local approximation (Acheson 1978). 

To close linearized equations, we require another equation obtained 
from the thermodynamic processes. The chemical equilibrium is realized in PNSs, and thus 
the density is a function of pressure, temperature, and lepton fraction. 
In the Boussinesq approximation, pressure perturbations are 
negligible because fluid elements are assumed to be in the pressure 
equilibrium with their surroundings. 
Thus the density and entropy perturbations can be expressed 
in terms of the perturbations of temperature and lepton fraction: 
\begin{equation} 
\delta \rho =  - \rho\Big(\alpha \displaystyle{\frac{\delta T}{T} 
+ \beta \delta Y_L}\Big) \;, \label{eq21} 
\end{equation}
\begin{equation}
\delta s = m_B c_p \dsf{\delta T}{T} + \zeta \delta Y_L \;, \label{eq22}
\end{equation}
where $m_B$ is the baryon mass, $\alpha=-(\partial\ln\rho /\partial\ln
 T)_{P,Y_L}$, $\beta = -(\partial\ln\rho/\partial Y_L)_{P,T}$
are the coefficients of thermal and chemical expansion, and 
$c_p = (T/m_B)(\partial s/\partial T)_{P,Y_L}$ is the specific heat at 
constant pressure. The sign of $\zeta = (\partial s/\partial Y_L)_{P,T}$ 
determines whether the leptonic gradient is stabilizing or destabilizing. 

Equations~(\ref{eq11}) and (\ref{eq12}) can be linearized and written in the form 
\begin{equation}
-i\omega_L \delta Y_L + \delta \mbox{\boldmath $u$} \cdot \nabla Y_L = 0 \;, \label{eq23}
\end{equation}
\begin{equation}
-i\omega_T \delta T - \delta \mbox{\boldmath $u$} \cdot \nabla \Phi_T
= \displaystyle{\frac{\zeta T}{\rho c_p}}b_L k^2 \delta Y_L \;, \label{eq24}
\end{equation}
where 
\begin{equation}
\omega_T = \omega + i \chi k^2 \;, \ \ 
\omega_L = \omega + i \xi k^2 \;, \label{eq25}
\end{equation}
\begin{equation}
\nabla \Phi_T = - \displaystyle{\frac{T}{m_B c_p}} (\nabla s - \zeta \nabla Y_L )
= \Big(\displaystyle{\frac{\partial T}{\partial P}}\Big)_{s,Y_L} \nabla P - \nabla T \;. 
\label{eq26}
\end{equation}
We introduce the heat diffusivity $\chi = a_T/\rho c_p $ and 
the chemical diffusivity $\xi = b_L/n$. 
Equation~(\ref{eq26}) represents the subadiabatic temperature gradient. 
Substituting equations~(\ref{eq23}) and (\ref{eq24}) into equation~(\ref{eq21}), 
we obtain the linearized equation, 
\begin{equation}
i\omega_T\omega_L \displaystyle{\frac{\delta \rho}{\rho}} 
= \alpha \displaystyle{\frac{\omega_L}{T}} (\delta \mbox{\boldmath $u$}\cdot\nabla \Phi_T)
-\beta [\omega + i ( 1 + \psi ) \chi k^2] (\delta \mbox{\boldmath $u$}\cdot \nabla Y_L) \;, \label{eq27}
\end{equation}
where 
\begin{equation}
\psi \equiv  \displaystyle{\frac{3}{\pi^2}}\displaystyle{\frac{\mu_{\nu_e}}{k_B T}} 
\Big( \displaystyle{\frac{\partial\ln\mu_{\nu_e}}{\partial \ln T}} \Big)_{s,P} 
\Big( \displaystyle{\frac{\partial \ s/k_B }{\partial Y_L}} \Big)_{T,P} \nonumber \;. \label{eq28}\\
\end{equation}
Assuming the nondegenerate neutrino and anti-neutrinos in degenerate 
nuclear matters, $\psi$ is considered to be sufficiently smaller than unity (Burrows \& Lattimer 1986). 
Thus, equation~(\ref{eq27}) is rewritten as  
\begin{equation}
i\omega_T\omega_L \displaystyle{\frac{\delta \rho}{\rho}} 
= \alpha \displaystyle{\frac{\omega_L}{T}} (\delta \mbox{\boldmath $u$}\cdot\nabla \Phi_T)
-\beta \omega_T (\delta \mbox{\boldmath $u$}\cdot \nabla Y_L) \;. \label{eq29}
\end{equation}
Eliminating perturbed quantities in eight linearized equations
[eqs. (\ref{eq14}) -- (\ref{eq20}) and (\ref{eq29})], we obtain the
following dispersion equation: 
$$
\displaystyle{\frac{k^2}{k^2_z}}\omega_D^4 - \displaystyle{\frac{k^2}{k^2_z}}\Big[
\displaystyle{\frac{\omega_{\eta}}{\omega_T}} N_{Te}^2 + 
\displaystyle{\frac{\omega_{\eta}}{\omega_L}} N_{Le}^2 \Big] \omega^2_D 
- \kappa^2 \omega^2_R \Big[ 1 + \displaystyle{\frac{2 \omega_A^2}{\kappa^2}}
\displaystyle{\frac{\omega_D^2}{\omega_R^2}} \Big]  
$$
\begin{equation}
-4\Omega^2(\mbox{\boldmath $k$}\cdot\mbox{\boldmath $v_A$})^2 
\Big[ 1 + \Big( \displaystyle{\frac{\omega_A}{\Omega}} \Big)^2+ 
\displaystyle{\frac{4-q}{2}} 
\Big( \displaystyle{\frac{\omega_A}{\Omega}} \Big)
\Big( \displaystyle{\frac{\omega_{\eta}}{\mbox{\boldmath $k$}\cdot\mbox{\boldmath $v_A$}}} \Big) 
+ \displaystyle{\frac{q}{2}} 
\Big( \displaystyle{\frac{\omega_A}{\Omega}} \Big)
\Big( \displaystyle{\frac{\omega_{\nu}}{\mbox{\boldmath $k$}\cdot\mbox{\boldmath $v_A$}}} \Big) \Big] = 0  \label{eq30}\;,
\end{equation}
where
$$
\omega^2_D = \omega_{\nu}\omega_{\eta} - (\mbox{\boldmath $k$}\cdot\mbox{\boldmath $v_A$})^2\;, \ \ \ 
\omega^2_R = \omega^2_{\eta} - (\mbox{\boldmath $k$}\cdot\mbox{\boldmath $v_A$})^2\;,
$$
$$
\mbox{\boldmath $v_A$} = \displaystyle{\frac{\mbox{\boldmath $B$}}{\sqrt{4\pi\rho}}} = (0,\ V_{A\phi},\ V_{Az})\;, \ \ \
\kappa^2 = \displaystyle{\frac{1}{r^3}}\displaystyle{\frac{\partial}{\partial r}}(r^4\Omega^2) \;,
$$
$$
\omega^2_A = \displaystyle{\frac{V_{A\phi}^2}{r^2}} = \displaystyle{\frac{B^2_{\phi}}{4\pi\rho r^2}} \;, \ \ \ 
\omega^2_{Az} = \displaystyle{\frac{V_{Az}^2}{r^2}}= \displaystyle{\frac{B^2_z}{4\pi\rho r^2}} \;, \ \ \
$$
$$
N_{Te}^2 = - \displaystyle{\frac{k^2_z}{k^2}}\displaystyle{\frac{\alpha}{\rho T}} (DP)D\Phi_T \;, \ \ \ 
N_{Le}^2 = \displaystyle{\frac{k^2_z}{k^2}}\displaystyle{\frac{\beta}{\rho}} (DP) DY_L \;, 
$$
$$
D \equiv \Big(\displaystyle{\frac{\partial }{\partial r}} - \displaystyle{\frac{k_r}{k_z}} 
\displaystyle{\frac{\partial }{\partial z}}\Big) \;, \ \ \ 
q \equiv - \displaystyle{\frac{d \ln \Omega}{d \ln r}}
$$
Here, $\mbox{\boldmath $v_A$}$ is the Alf\'ven velocity, $\kappa$ is the epicyclic frequency, 
$\omega_A$ is the azimuthal Alfv\'en frequency, 
$\omega_{Az}$ is the vertical Alfv\'en frequency, 
$N_{Te}$ is the effective thermal buoyancy frequency, 
and $N_{Le}$ is the leptonic buoyancy frequency. 
The shear parameter $q $ denotes the rotational configuration of the systems. 

Since the dispersion equation (\ref{eq30}) has a very complex form, 
it is difficult to treat for what it is. 
Therefore, we simplify it using reasonable approximations. 
When we focus on the MRI modes with $\omega \sim \mbox{\boldmath $k$}\cdot \mbox{\boldmath $v_A$}$ 
and adopt the weak magnetic field approximation $\Omega \gg \omega_A $, 
equation~(\ref{eq30}) can be described as 
\begin{equation}
\displaystyle{\frac{k^2}{k_z^2}}\omega_D^4 - 
\displaystyle{\frac{k^2}{k_z^2}}\Big[ \displaystyle{\frac{\omega_\eta }{\omega_T}}N_{Te}^2 + 
\displaystyle{\frac{\omega_\eta}{\omega_L}}N_{Le}^2 \Big] \omega_D^2 
- \kappa^2 \omega^2_R 
-4(\mbox{\boldmath $k$}\cdot\mbox{\boldmath $v_A$})^2\Omega^2 = 0 \;. \label{eq31}
\end{equation}
It is stressed that both axisymmetric and nonaxisymmetric MRI modes are included in 
this dispersion equation. In the diffusionless ($\nu=\eta=\chi=\xi=0$) and 
pure toroidal ($B_z=0$) limits, equation~(\ref{eq31}) is identical to 
the result of Paper I. Taking chemically homogenious ($ \nabla Y_L =0$) 
and axisymmetric ($m=0$) limits, it reduces to the dispersion equation 
derived by Menou et al.2004.

\subsection{Stability Criterion}
The nature of instabilities in PNSs has been considered by a number of 
authors (Bruenn \& Dinneva 1996; Mezzacappa et al.1998; Miralles et al.2000,2002,2004). 
We make a detailed comparison of our dispersion equation 
with previous studies and derive the stability criterion for the MRI in the neutrino loaded PNSs. 

Taking the diffusionless and hydrodynamic ($\mbox{\boldmath $B$} = 0$) limits, 
equation~(\ref{eq31}) is identical to the result of Miralles et al.2004, in which 
the stability conditions for the differentialy rotating PNSs are obtained. 
In the non-rotating limit ($\Omega = 0$), equation~(\ref{eq31}) reduces to 
the results of Miralles et al.2002, in which they derive general stability criteria 
for the convective instability, taking dissipative processes 
into account such as neutrino transport, viscosity, and resistivity. 
Thus, the both semiconvective and neutron-finger unstable modes are obviously involved 
in our dispersion equation (Bruenn \& Dinneva 1996; Miralles et al.2000). 
It is, however, noticed that we neglect the influence of the chemical inhomogeneity 
on the heat diffusion and the thermal inhomogeneity on the chemical diffusion for simplicity. 

Although the linear and nonlinear growth of 
semiconvection and neutron-finger instability would be a quite important issue 
in the evolution process of PNSs, it is beyond the scope of this paper. 
We now have an interest with the effects of multiple diffuions on the growth of MRI. 
Hence, we show the general stability criterion for the MRI in multi-diffusive systems, 
which is derived from equation~(\ref{eq31}) with the same procedure as in Urpin 2006. 
When we focus on relatively larger wavelengths ($\lambda \gg \lambda_c \equiv 2\pi\sqrt{\nu\eta}/V_{Az}$), 
the stability criterion for the MRI can be written as 
\begin{equation}
\displaystyle{\frac{\eta}{\chi}} N_T^2 + \displaystyle{\frac{\eta}{\xi}} N_L^2 
+ \displaystyle{\frac{\rm{d}\Omega^2}{\rm{d}\ln r}} > 0 \label{eq32} \;,
\end{equation}
where the thermal buoyancy frequency $N_T$ and the leptonic buoyancy frequency $N_L$ are defined by 
\begin{equation}
N_T^2 = - \displaystyle{\frac{\alpha}{\rho T}}(\nabla P)\cdot\nabla\Psi_T \;, \ \ \ 
N_L^2 = \displaystyle{\frac{\beta}{\rho}}(\nabla P)\cdot\nabla Y_L \label{eq33} \;.
\end{equation}
Equation~(\ref{eq32}) have a form similar to that derived in Urpin 2006 but with 
the leptonic gradient and the chemical diffusion. 
This criterion differs essentially from the standard stability criterion of the MRI 
(Balbus \& Hawley 1991,1994). Considering the diffusive coefficients 
which are typical in PNSs ($\chi ,\xi , \gg \eta$: see \S~4.1), 
the stabilizing effect of the stratification is much reduced 
in our criterion. Here we notice that the viscous effect is not appeared in above criterion. 
However, the growth of MRI can be suppressed by the viscous damping (see \S~3.2). 
Thus we should consider not only the stability criterion but also the unstable growth rate 
when we discuss the characteristics of the MRI in multi-diffusive situations. 

\section{GENERAL FEATURES OF THE DISPERSION EQUATION}
We solve the dispersion equation~(\ref{eq31}) numerically 
and examine the characteristics of MRI in 
multi-diffusive situations. Note that, within the local approximation, 
the dispersion equation~(\ref{eq31}) is valid in the case 
$\Omega \gtrsim \mbox{\boldmath $k$}\cdot\mbox{\boldmath $v_A$} \gg \omega_A$. 
As described above, both axisymmetic and nonaxisymmetic MRI modes 
are included in our dispersion equation. 
General features of MRI in such complicated situations 
are not fully investigated in previous studies. 
Our analysis could be applicable to not only the PNSs but also 
the other astrophysical objects, such as stellar interiors and accretion disks. 

The effective thermal and leptonic buoyancy frequencies, 
$N_{Te}$ and $N_{Le}$, denote the size of the restoring force due to the stable 
stratification. Qualitative effect of the leptonic buoyancy is the same as thermal one. 
Thus we consider only the thermal buoyancy frequency as the measure 
of stabilizing effects due to the stratification for a while. 
In PNSs, the toroidal magnetic fields would dominate over the poloidal one. 
In the following, we fix the ratio of the toroidal to poloidal Alfv\'en frequencies 
as $\omega_{A}/\omega_{Az} = 10^4$. The ratio of the angular velocity 
to the toroidal Alfv\'en frequency is assumed to be $\Omega/\omega_{A} = 100$. 
These are about their typical values in rotating PNSs (see \S~4). 
Since we have an interest to the systems with differential rotation, 
the shear parameter $q=1.5$ is adopted in this section. 
Furthermore, we restrict our analysis to the modes with zero radial wavenumber ($k_r=0$) which 
correspond to the fastest growing branch of the axisymmetric and nonaxisymmetric MRI. 

\subsection{Diffusionless Limit}
\begin{figure}
\epsscale{.80}
\plotone{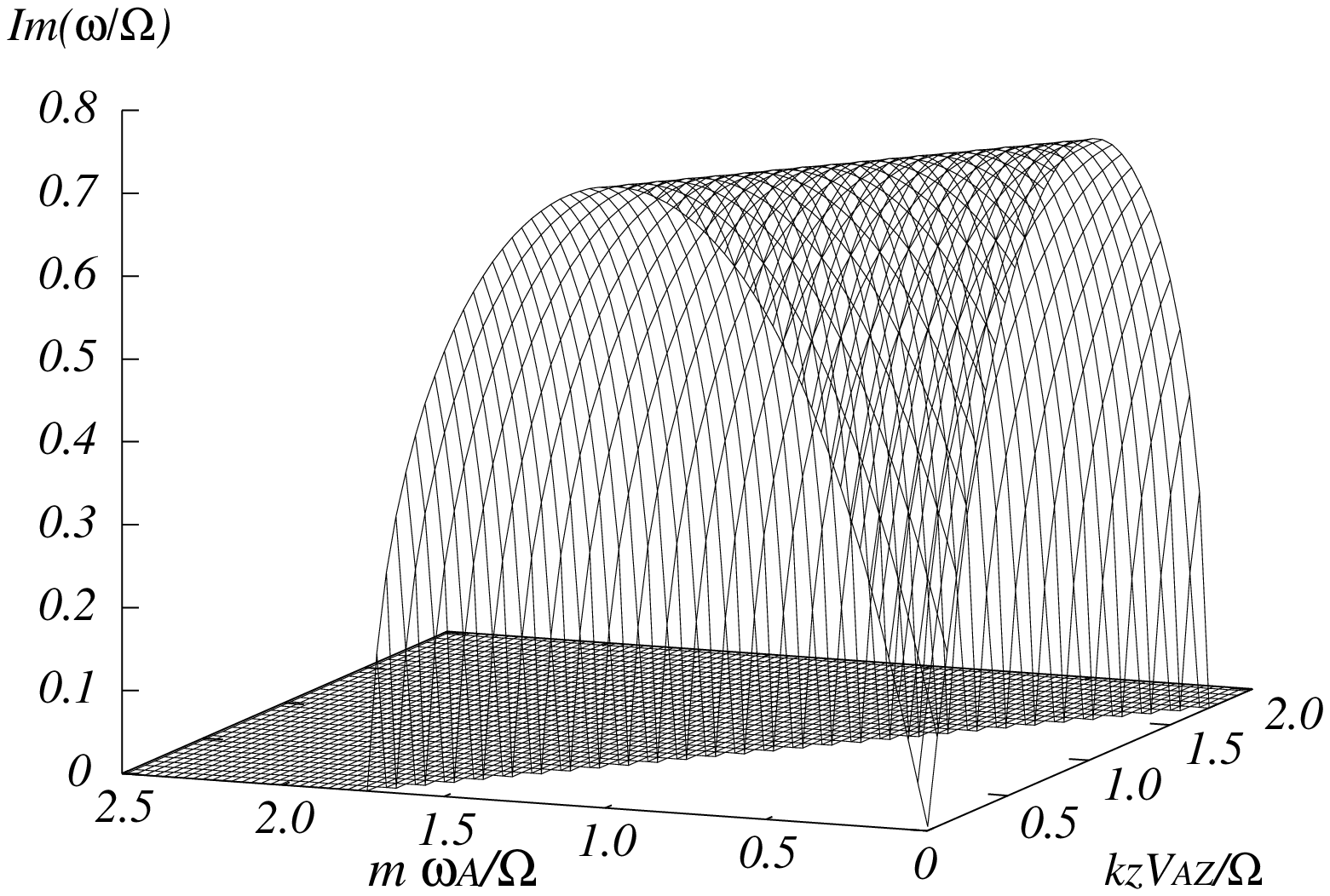}
\caption{Unstable growth rate $\Im (\omega)$ in the 
diffusionless case is plotted as a function 
the azimuthal wavenumber $m$ and the vertical wavenumber $k_z$. 
For simplicity, we select the case of zero radial wavenumber $k_r =0$. 
The growth rate is normalized by the 
angular velocity $\Omega$, and the wavenumbers are normalized by  
$\omega_A/\Omega$ and $V_{Az}/\Omega$, respectively. 
In this figure, we assume the shear parameter $q=1.5$, 
the effective thermal and leptonic buoyancy frequencies $N_{Te} =N_{Le} = 0$. 
Quadruple diffusivities are assumed to be zero ($\chi=\xi=\nu=\eta=0$).}
\label{fig1}
\end{figure}

We show the features of the dispersion relation in the diffusionless limit 
($\nu=\eta=\chi=\xi=0$). When the imaginary part of $\omega$ is positive, 
the amplitude of the perturbation will grow exponentially. 
Figure~\ref{fig1} shows a three-dimensional plot of the unstable growth rate 
as a function of the vertical and azimuthal wavenumbers. 
Stabilizing effects due to the stratification are ignored here ($N_{Te} =0$). 
The growth rate is normalized by angular velocity $\Omega$. 
Both axisymmetric and nonaxisymmetric modes can be unstable for the MRI 
in this situation. For the case with $q=1.5$, 
the maximum growth rate of axisymmetric modes is 
$0.75\Omega$ at $k_z = k_{\rm{max}} = 0.97\Omega/V_{Az}$. 
The cutoff wavenumber of axisymmetric modes 
depend on the strength of the vertical field, which is given by 
$k_{\rm{crit}} = 1.73\Omega/V_{Az}$ (Balbus \& Hawley 1991). 
The maximum growth rate of nonaxisymmetric modes ($m\ne 0$) is $0.75\Omega $ at 
$m= m_{\rm{max}} = 0.97\Omega/\omega_A$. 
The cutoff wavenumber is determined by the strength of 
the toroidal field and given by $m_{\rm{crit}} =1.73\Omega/\omega_A$ (see Paper I). 

The stable stratification acts on the MRI 
as the negative buoyancy and reduces its growth rate. 
The amplitude of the restoring force is denoted by the size of $N_{Te}$. 
The stabilizing effect due to the stratification on the axisymmetric ($m=0$) and 
nonaxisymmetric ($m=m_{\rm{max}}$) MRI are depicted in Figure~\ref{fig2}. 
The normalized growth rate is given by a function of the vertical wavenumber 
for the cases with different stabilizing parameters $N_{Te}/\Omega = 0, 0.5,1$ and $2$. 
The horizontal axes are measured by the logarithmic scale. 
As seen from these figures, the growth rate of MRI decreases 
as $N_{Te}/\Omega$ increases. When the stabilizing parameter 
exceeds a critical value ($N_{Te} /\Omega \gtrsim 2$), 
the growth of MRI is completely suppressed. 
These results are generally approved in spite of the radial wavenumbers. 

\subsection{Effects of Quadruple-Diffusions}
Our main interest in this paper is to understand the effects of 
multiple diffusions on the growth of MRI. 
In general, the viscous, heat, and chemical diffusions have a major impact on 
the growth of hydorodynamic and magnetohydrodynamic instabilities (Bruenn \& Dinneva 1996; 
Miralles et al. 2000, 2002; Bruenn et al. 2004). 
The ohmic dissipation could also affect the growth of MRI 
(Jin 1996; Sano \& Miyama 1999). 
We examine the single and multi-diffusive effects on the growth of MRI. 
We restrict our discussion to the case with zero radial wavenumber ($k_r = 0$). 
For further simplicity, we fix the azimuthal wavenumber as 
$m = m_{\rm{max}} $, which corresponds to 
the fastest growing mode of nonaxisymmetric MRI (see \S~3.1). 

The dispersion relation is characterized by the normalized diffusivities defined as 
\begin{equation}
Pe \equiv \displaystyle{\frac{V_{Az}^2}{\chi\Omega}}\;, \ \ 
Re \equiv \displaystyle{\frac{V_{Az}^2}{\nu\Omega}} \;, \ \ 
Re_M \equiv \displaystyle{\frac{V_{Az}^2}{\eta\Omega}} \;, \label{eq40}
\end{equation}
where $Pe$ is the Peclet number, $Re$ is the Reynolds number, 
and $Re_M$ is the magnetic Reynolds number. 
Here, we select the vertical Alfv\'en speed as the typical velocity and 
$V_{Az}/\Omega$ as the typical lengthscale. 
The effect of the chemical diffusion is quite similar to that of the heat diffusion, 
and thus it is ignored for a while. 
We now consider the case with a strong toroidal field $\omega_A/\omega_{Az} = 10^4$. 
The characteristic length scale of nonaxisymmetric MRI is thus 
much longer than that of axisymmetric MRI. 
Therefore, axisymmetric MRI is affected by the diffusion processes 
more severely than nonaxisymmetric MRI. 

\subsubsection{Single Diffusive Effects} 
It would be useful to describe the effects of each diffusions on the MRI separately. 
Figures~\ref{fig3}a, \ref{fig3}b and \ref{fig3}c show the dispersion relation taking account of the 
heat, viscous, and magnetic diffusivities respectively. 
The horizontal axes of these figures are vertical wavenumber in the logarithmic scale. 

Figure~\ref{fig3}a demonstrates the effect of the heat diffusion. 
The other diffusive coefficients are assumed to be $\nu=\eta=0$. 
The heat diffusion can reduce the stabilizing effect due to the stratification 
and revive the unstable growth of MRI. 
As described in \S~3.1, the growth of MRI is 
suppressed by the restoring force of negative buoyancy. 
However, even with a strong stratification ($N_{Te}/\Omega \gtrsim 2$), 
MRI can grow if the heat diffusion is efficient. 
This is because the heat diffusion induces the heat exchange 
between the fluid element and the surroundings, which eliminates 
the entropy difference between them. 
Thus it reduces the stabilizing effects caused by the entropy gradients 
and the growth of MRI is revived. 

The typical wavelength $\lambda_{\chi}$ at which the heat diffusion can 
affect the buoyant oscillation is 
estimated from a simple relation, $\chi k^2 \sim N_{Te}$, that is 
\begin{equation}
\lambda_\chi = 2\pi \sqrt{\displaystyle{\frac{\chi}{N_{Te}}}} \;. \label{eq41} 
\end{equation}
The heat diffusion affects the modes with shorter wavelengths than $\lambda_\chi$. 
On the other hand, the cutoff wavelength of MRI 
is given by $\lambda_{\rm{MRI}} \sim 2\pi V_{Az}/\Omega$. 
Magnetic tensions stabilize the modes with longer wavelengths than $\lambda_{\rm{MRI}}$. 
Thus we can derive the range of poloidal wavelength which involve fast growing MRI modes 
in the case with heat diffusion and stable stratification, that is 
$ \lambda_{\rm{MRI}} \lesssim \lambda \lesssim \lambda_\chi $, in other words, 
\begin{equation}
1 \lesssim \bar{\lambda} \lesssim \bar{\lambda}_\chi \equiv \sqrt{\frac{\Omega}{N_{Te} Pe}}
\;, \label{eq42}
\end{equation}
where $\bar{\lambda} \equiv \lambda/\lambda_{\rm{MRI}}$. 
Here we define the fast growing MRI modes as the modes whose growth rate is 
of the order of $q\Omega$. 
For the case with $N_{Te}/\Omega =2$, 
the condition for reviving the MRI is $P_e \lesssim 0.5$. 
This is consistent with the result shown in Figure~\ref{fig3}a. 
In the strongly stratified regions ($ N_{Te}/\Omega \gg 1$), 
such as the outer layer of PNSs and the radiative core of the Sun, 
the condition $P_e \ll 1$ is needed for the growth of MRI. 

Figure~\ref{fig3}b shows the viscous damping of the growth of MRI. 
Here $N_{Te} $ is assumed to be zero and the 
heat and magnetic diffusivities are neglected ($\chi = \eta = 0$). 
The effect of the kinetic viscosity is characterized by the Reynolds number $Re$. 
The Reynolds number is unity when the viscous damping rate 
is comparable to the growth rate of MRI, $\nu k^2 \sim \Omega$. 
The growth of MRI is suppressed by the viscosity when 
$Re \lesssim 1$ as shown in Figure~\ref{fig3}b. The critical wavelength is 
approximately given by 
\begin{equation} 
\bar{\lambda}_\nu \equiv \displaystyle{\frac{\lambda_\nu}{\lambda_{\rm{MRI}}}} = 
\sqrt{\displaystyle{\frac{\nu \Omega}{V_{Az}^2}}} = \sqrt{\displaystyle{\frac{1}{Re}}} \;, \label{eq43}
\end{equation} 
where $\lambda_\nu \equiv 2\pi\sqrt{\nu/\Omega}$. 
We notice that the growth of MRI is suppressed but never stabilized by 
the viscous damping. This means that the stability condition of MRI is 
not affected by the viscous dissipation, which is discussed in \S~2.3. 
Thus we should consider both the growth rate and the stability condition 
when we discuss the characteristics of MRI in multi-diffusive systems. 

Ohmic dissipative effects are shown in Figure~\ref{fig3}c. 
The conditions $N_{Te}=0$ and $\chi = \nu =0$ are 
assumed in this figure. 
As seen from this figure, the effect of ohmic dissipation is 
similar to that of viscous dissipation. 
However, the ohmic dissipation can stabilize the MRI when 
the following condition is satisfied, 
\begin{equation}
\bar{\lambda}  \lesssim \bar{\lambda}_\eta \equiv \displaystyle{\frac{\lambda_\eta}{\lambda_{\rm{MRI}}}}  
= \sqrt{\displaystyle{\frac{1}{Re_M}}} \;, \label{eq44}
\end{equation}
where $\lambda_\eta \equiv 2\pi\sqrt{\eta/\Omega}$. 
Thus the stability condition of MRI depends on the magnetic diffusivity (see,\S~2.3). 
The ohmic dissipation becomes effective when $Re_M \lesssim 1$ is satisfied. 
This is also consistent with the Figure~\ref{fig3}c. 

\subsubsection{Double Diffusive Effects}
Multi-diffusive processes often affect the stability of astrophysical objects. 
We investigate the double diffusive effect on the growth of MRI 
focusing on typical two cases. 
One is the double diffusive system with $\chi \gg \nu $ (PNS-type system), 
and the other is that with $\chi \gg \eta $ (SOLAR-type system). 
As described later, heat and viscous diffusions caused by the neutrino radiation 
dominate over the magnetic diffusion in PNSs ($\chi \gg \nu \gg \eta$). 
On the other hand, the condition $\chi \gg \eta \gg \nu$ is 
satisfied in the solar radiative zone. 
We restrict our discussion to the cases with zero radial wavenumber 
and the azimuthal wavenumber $m= m_{\rm{max}} $. 

Figure~\ref{fig4}a shows the double diffusive effect on the growth of MRI 
in the PNS-type system. Vertical and horizontal axes are the same as Figure~\ref{fig3}. 
The thick dashed-line shows the growth rate in the diffusionless 
and isentropic limits. The stabilization of MRI due to the stable stratification 
is depicted by the thick dotted-line ($N_{Te}/\Omega =2$). 
The solid line shows the double diffusive effect of heat and viscous diffusions 
in the PNS-type system. We consider the case with $\chi \gg \nu $, 
so that $Pe = 10^{-4}$ and $Re = 0.01$ are assumed here. 
Then the critical wavelengths are $\bar{\lambda}_\chi = 10^2$ 
and $\bar{\lambda}_\nu = 10$, respectively. 

From this figure, it is found that the stabilizing effect 
is reduced and the growth of MRI is revived for the modes 
with the poloidal wavelengths shorter than $\lambda_\chi $. 
On the other hand, the unstable growth is suppressed by the viscous damping 
for the modes with the poloidal wavelengths shorter than $\lambda_\nu$. 
Therefore, in the PNS-type system, fast growing modes of MRI are limited to 
the following range; 
\begin{equation}
\max (\lambda_\nu,\lambda_{\rm{MRI}} ) \lesssim \lambda \lesssim \lambda_\chi \;. \label{eq45}
\end{equation} 
We show the relation between single and double diffusive cases. 
The thin dotted-line shows the single diffuisive case with the heat 
diffusion of $Pe = 10^{-4}$ and $N_{Te} = 2.0$.  
The case with only the viscous diffusion of $Re=0.01$ and $N_{Te}=0$ is described by 
the thin dashed-line. We find that the double diffusive case 
is obviously superposed by each single diffusive cases. 

Figure~\ref{fig4}b shows the double diffusive effect 
on the growth of MRI in the SOLAR-type system. 
Thick dashed and dotted-lines are the same as in Figure~\ref{fig4}a. 
The solid line shows the double diffusive effect in the SOLAR-type system 
($\chi \gg \eta $). When we assume $Pe = 10^{-4}$ and $Re_M = 0.01$, 
corresponding critical wavelengths are $\bar{\lambda}_\chi = 10^2 $ 
and $\bar{\lambda}_\eta = 10$, respectively. 
As seen from this figure, the unstable growth of the MRI are revived by the 
effect of heat diffusion the same as in the PNS-type system. 
However, the ohmic dissipation stablize 
the unstable growth of MRI in the regime $\lambda \lesssim \lambda_\eta$. 
Therefore, in the SOLAR-type system, fast growing modes of MRI are limited 
to the following range, 
\begin{equation}
\max (\lambda_\eta,\lambda_{\rm{MRI}}) \lesssim \lambda \lesssim \lambda_\chi \;. \label{eq46}
\end{equation} 
The SOLAR-type double diffusive case is also superposed by each single diffusive cases. 
The thin dashed-line shows the single diffuisive case with the 
thermal diffusion of $Pe = 10^{-4}$ and $N_{Te} = 2.0$.  
The case with only the ohmic dissipation of $Re_M=0.01$ and $N_{Te}=0$ is 
depicted by the thin dotted-line. 
It is found that the condition of faster growing modes 
are determined by single diffusive behaviors of the disprsion relation. 

Finally, we consider the general multi-diffusive systems 
($\chi\ne 0,\xi\ne0, \nu\ne 0, \eta\ne 0$). 
It is indicated from the double diffusive systems 
that the lower limit of the wavelength for 
fast growing modes is determined by the two dissipation processes, 
that is the viscous and ohmic dissipation. 
In addition, its upper limit is determined 
by the two relaxation processes, that is the heat and lepton diffusions. 
Thus, in the multi-diffusive system, fast growing modes of MRI are limited to
the following range: 
\begin{equation}
\max{( \lambda_\nu , \ \lambda_\eta , \ \lambda_{\rm{MRI}})} 
\lesssim \lambda \lesssim \min{ ( \lambda_\chi, \ \lambda_\xi ) }\;, \label{eq47}
\end{equation}
where $\lambda_\xi \equiv 2\pi \sqrt{\xi/N_{Le}}$. 
In stably stratified regions ($N_{Te}, N_{Le} > \Omega $) 
such as the envelope of PNSs and the solar radiative core, 
most of unstable MRI modes are drastically suppressed by 
the viscous damping, ohmic dissipation or negative buoyancy. 
Only the modes satisfying the condition given by equation~(\ref{eq47}) can grow significantly. 
The properties of the systems are determined by the fastest growing modes. 
It is, thus, important to know the typical lengthscale 
defined by the condition (\ref{eq47}). 

\subsubsection{Difference in Axisymmetric and Nonaxisymmetric MRI}
We have investigated the diffusive effects on the nonaxisymmetric MRI. 
Generally, fast growing modes of axisymmetric MRI are also limited 
to the range defined by equation (\ref{eq47}). 
However, the growth rate can differ among the axisymmetric 
and nonaxisymmetric MRI in multi-diffusive cases. 

\begin{figure}
\begin{center}
\begin{tabular}{c}
\scalebox{0.6}{\rotatebox{0}{\includegraphics{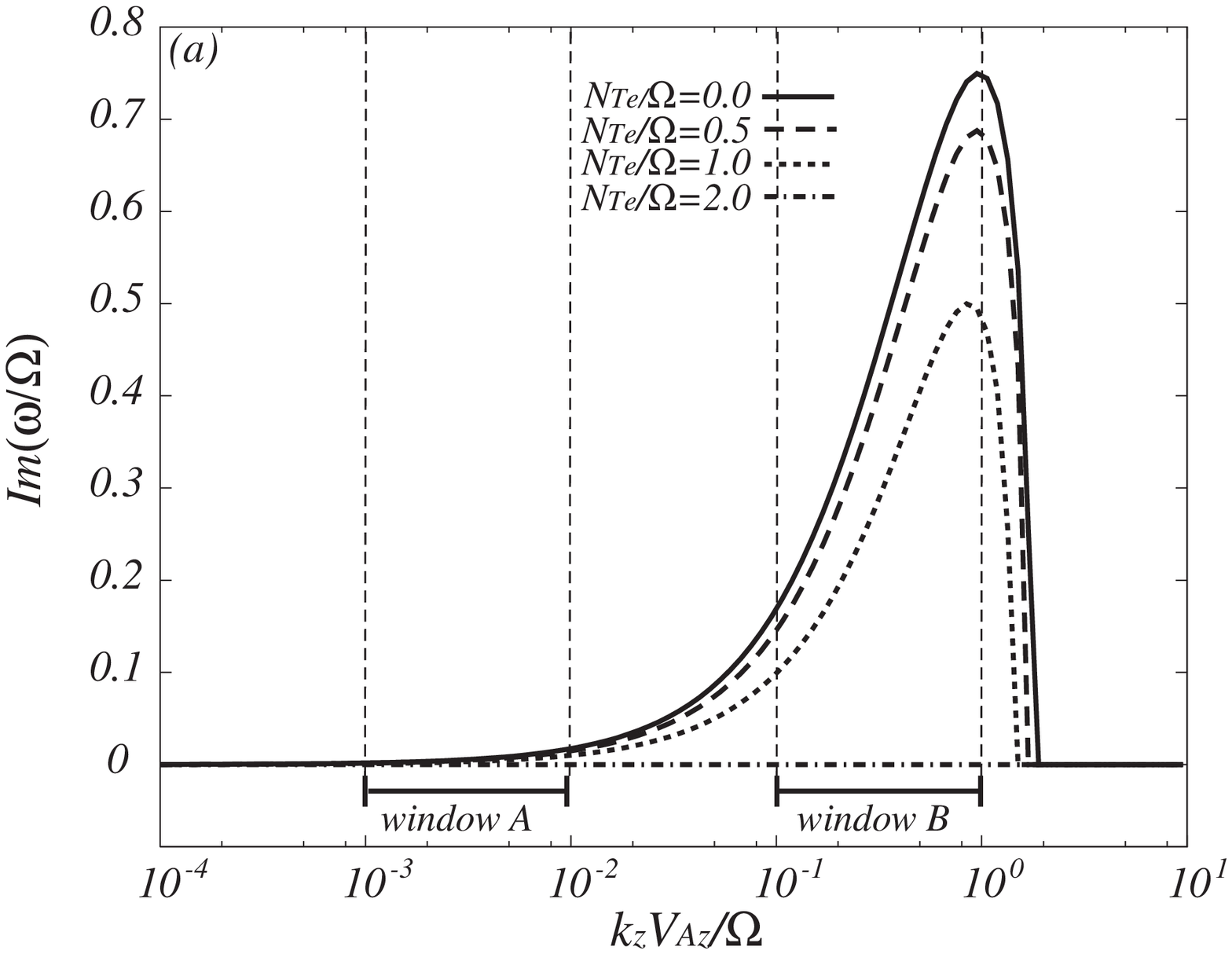}}} \\
\end{tabular}
\begin{tabular}{c}
\scalebox{0.6}{\rotatebox{0}{\includegraphics{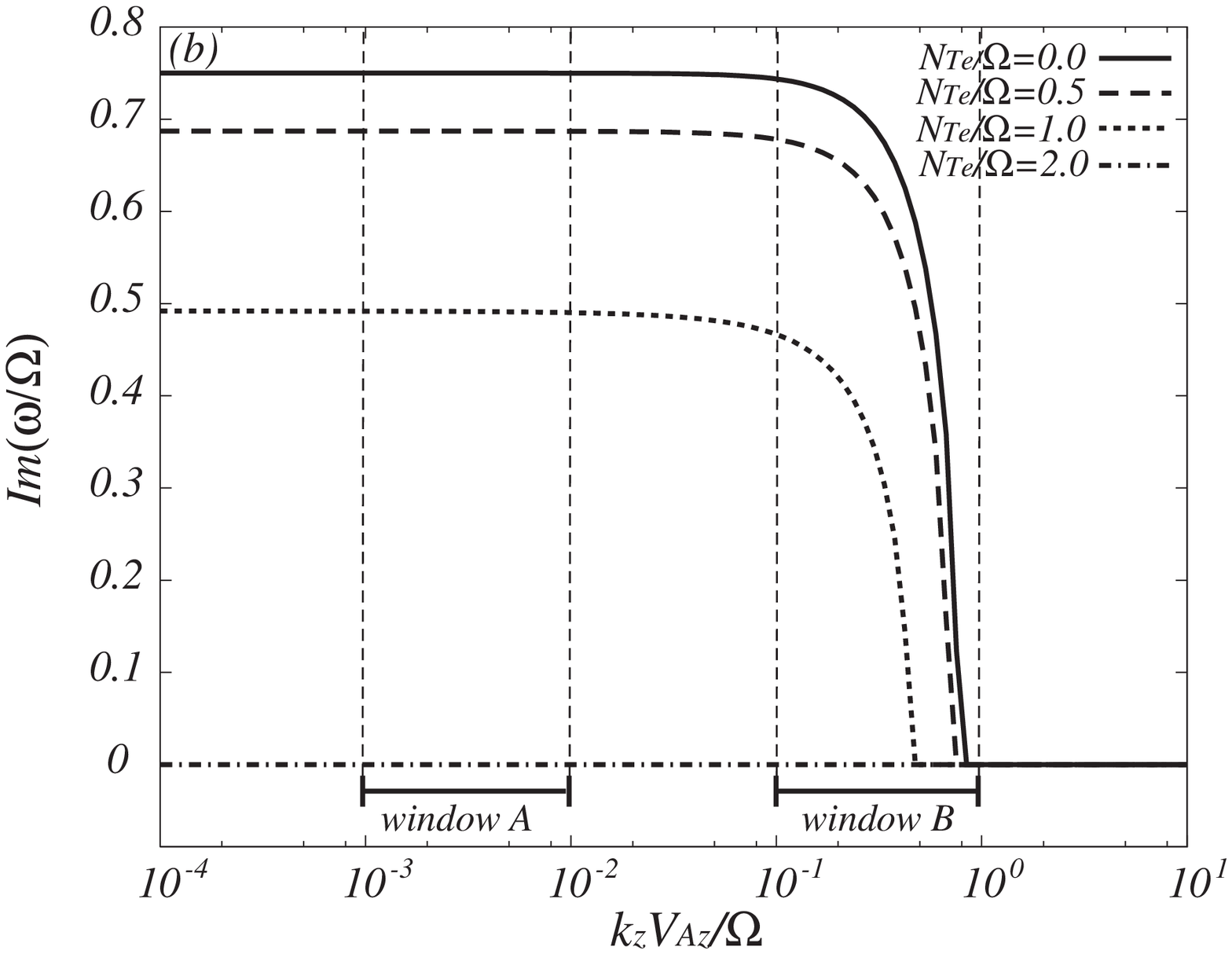}}} \\
\end{tabular}
\caption{Dependence of the growth rate on the stabilizing parameter $N_{Te}/\Omega$ 
for the case of (a) axisymmetric MRI ($m=0$), and (b) nonaxisymmetric MRI ($m=m_{\rm{max}}$). 
The growth rate is shown as a function of the normalized vertical wavenumber 
$k_z V_{Az}/\Omega$ for the case $N_{Te}/\Omega = 0, 0.5,1$, and $2$, respectively. 
Note that the horizontal axis is measured by the logarithmic scale. 
Model parameters are the same as those in Figure~\ref{fig1}. 
The growth rate of axisymmetric and nonaxisymmetric MRI decrease 
as the stabilizing parameter increases. 
When $N_{Te} \gtrsim 2\Omega$, the growth of the MRI is completely stabilized by 
the stable stratification. }
\label{fig2}
\end{center}
\end{figure}

\begin{figure}
\begin{center}
\begin{tabular}{c}
\scalebox{0.45}{\rotatebox{0}{\includegraphics{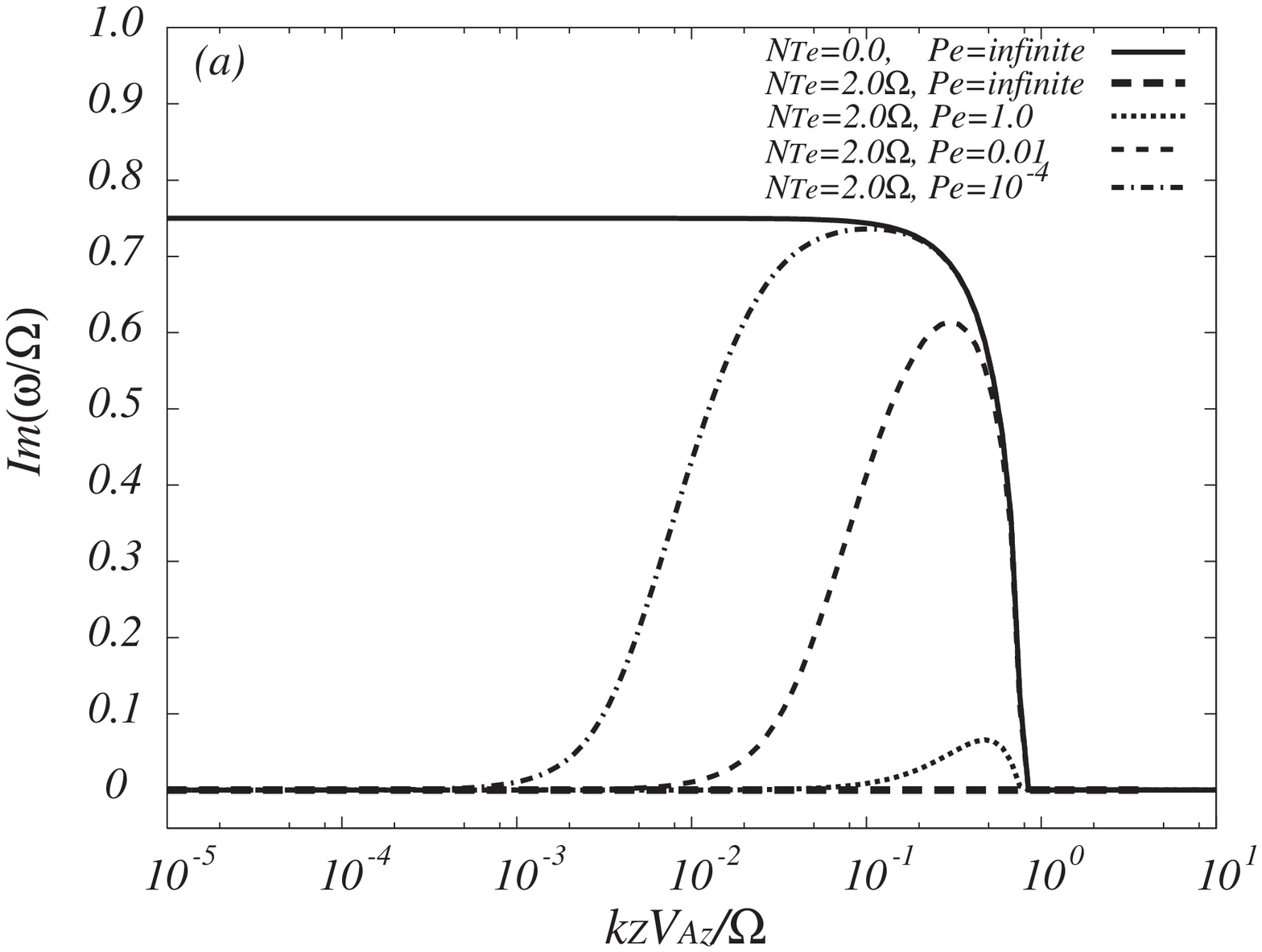}}} \\
\end{tabular}
\begin{tabular}{c}
\scalebox{0.45}{\rotatebox{0}{\includegraphics{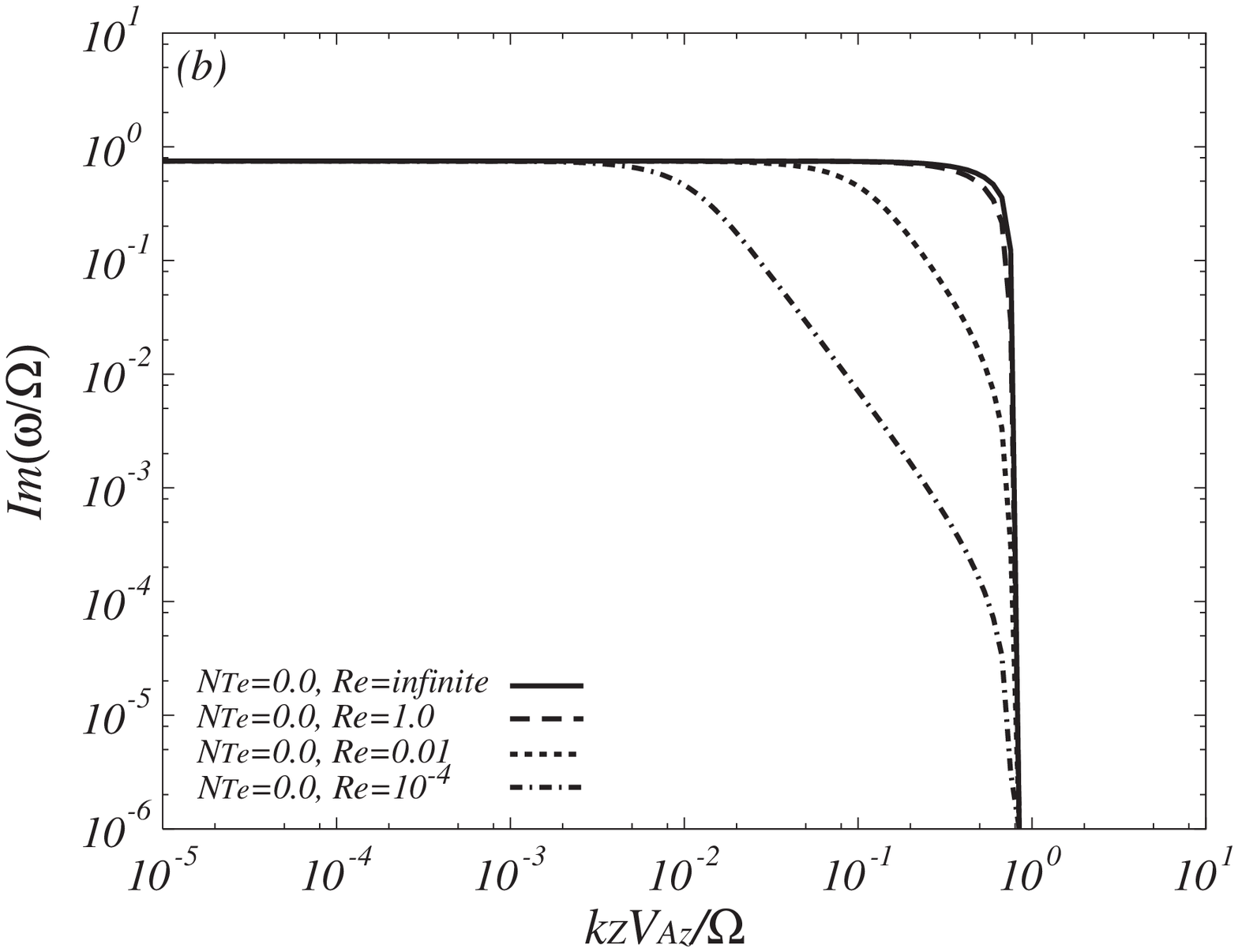}}} \\
\end{tabular}
\begin{tabular}{c}
\scalebox{0.45}{\rotatebox{0}{\includegraphics{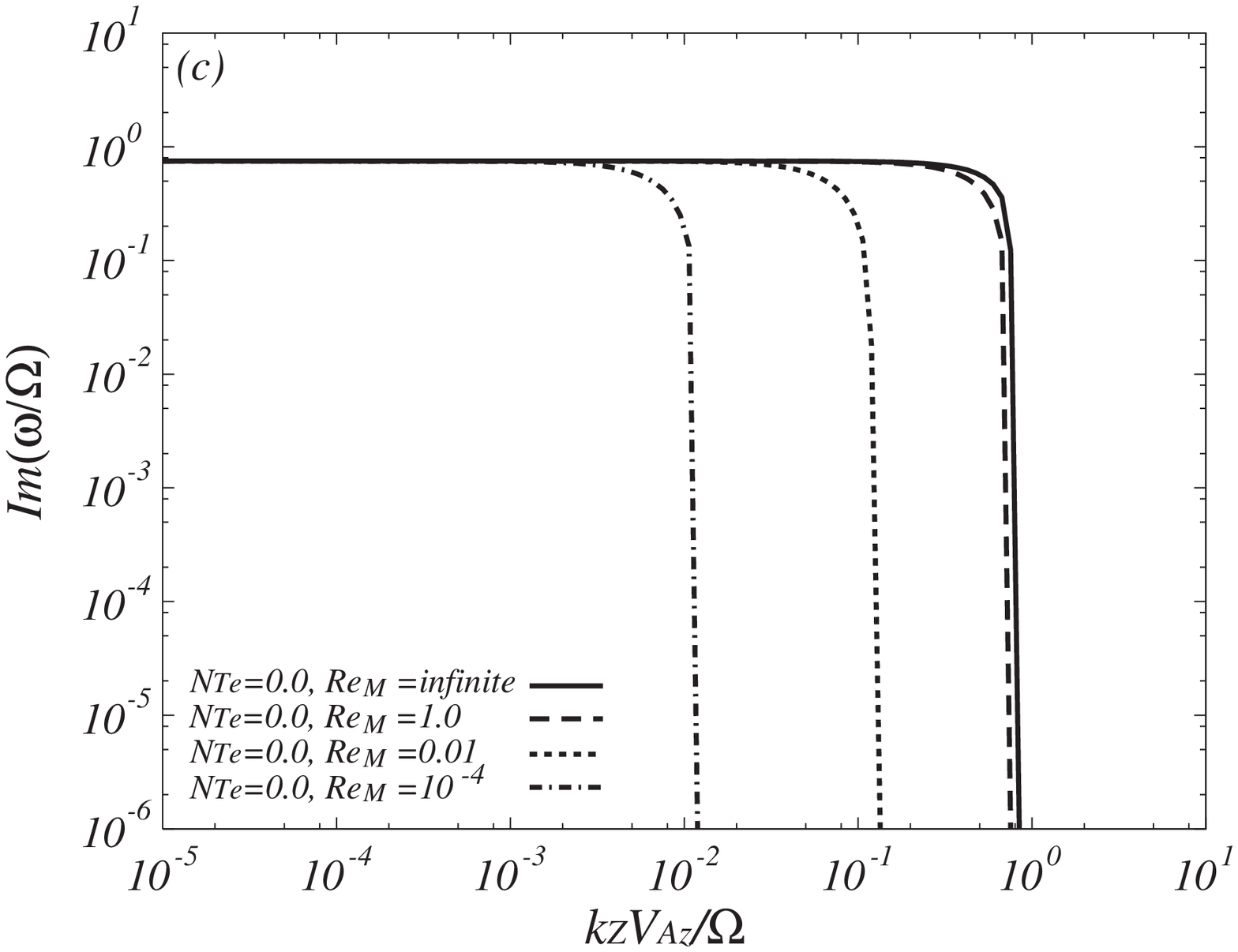}}} \\
\end{tabular}
\caption{Single diffusive effects of (a) heat, 
(b) viscous, and (c) magnetic diffusions on the growth of MRI. 
Normalized growth rates are shown as a function of 
the normalized vertical wavenumber $k_z V_{Az}/\Omega$. 
We use the Peclet number $Pe$, the Reynolds number $Re$ and the magnetic Reynolds number $Re_M$ 
as the measures of the effects of heat, viscous, and mangetic diffusivities respectively.  
For simplicity, we select the branch with $m=m_{\rm{max}}$ 
and $k_r=0$. This corresponds to the fastest growing branch of 
the nonaxisymmetric MRI. The other parameters are the same as in Figure~\ref{fig1}. 
}
\label{fig3}
\end{center}
\end{figure}

\begin{figure}
\begin{center}
\begin{tabular}{c}
\scalebox{0.65}{\rotatebox{0}{\includegraphics{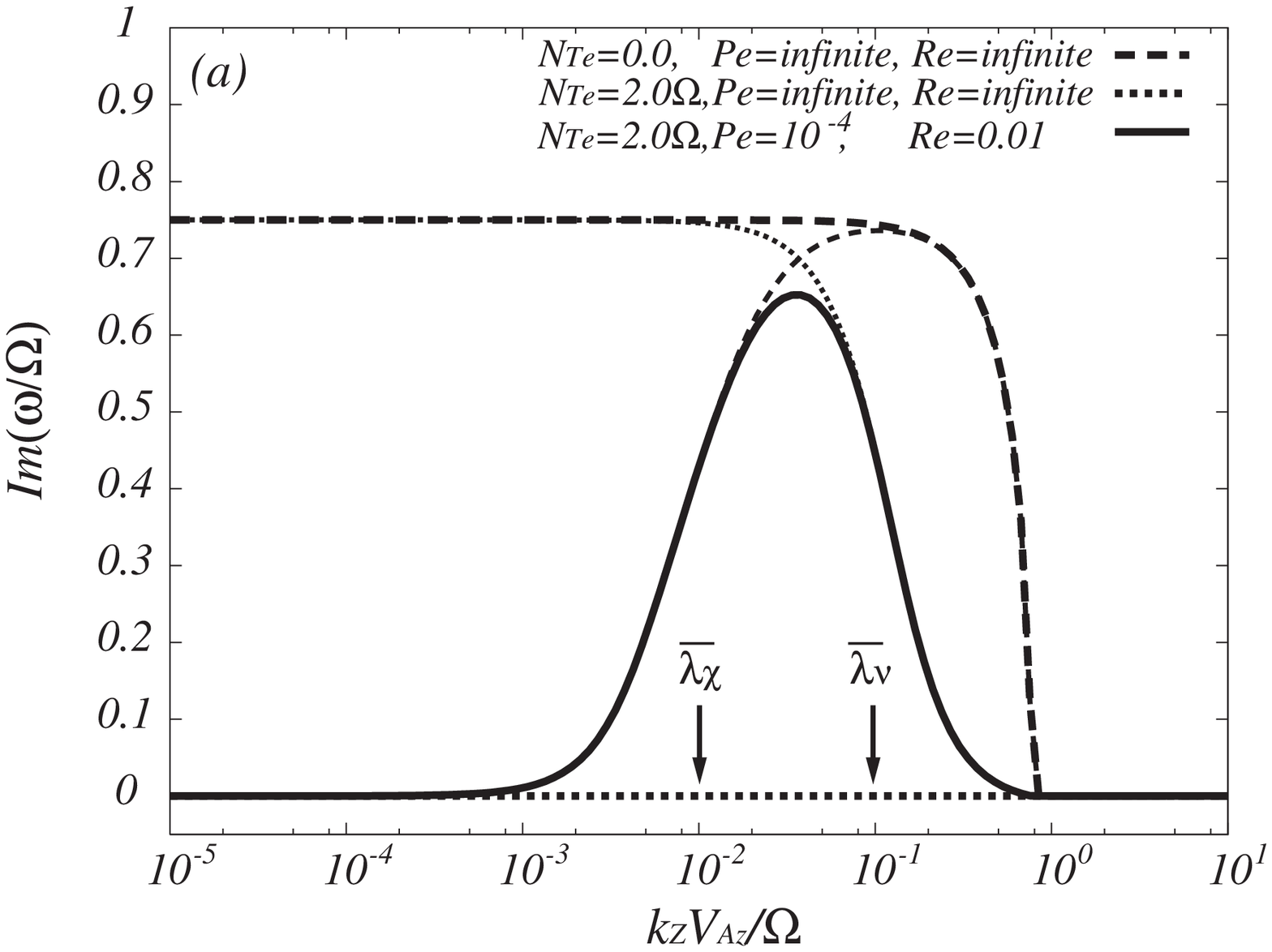}}} \\
\end{tabular}
\begin{tabular}{c}
\scalebox{0.65}{\rotatebox{0}{\includegraphics{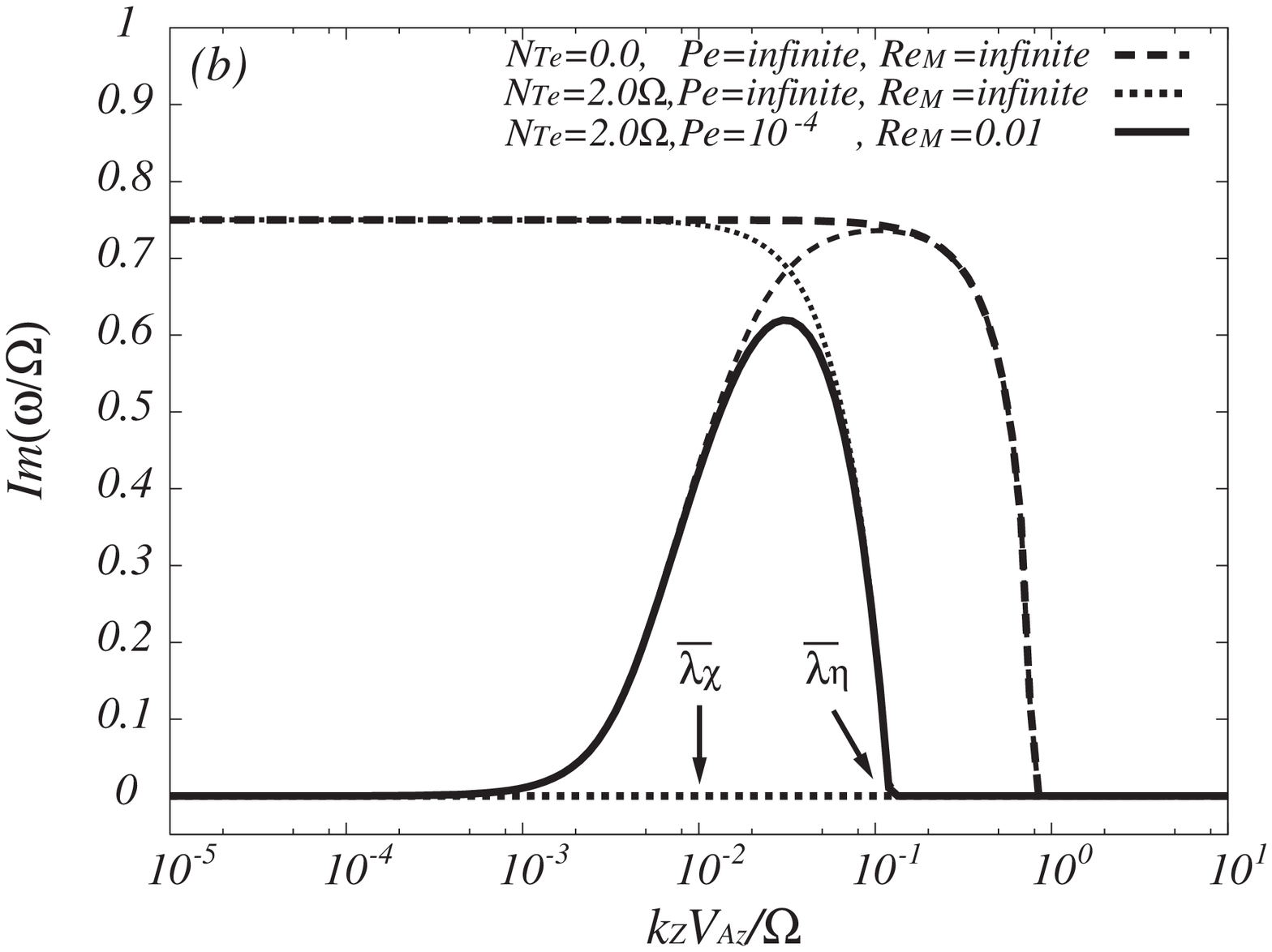}}} \\
\end{tabular}
\caption{Double diffusive effects in (a) the PNS-type systems and 
(b) the SOLAR-type systems. In the PNS-type systems, the heat and viscous diffusivities are 
sufficiently larger than the magnetic diffusivity ($\chi \gg \nu \gg \eta$). 
On the other hand, in the SOLAR-type systems, the heat and magnetic diffusivities are 
dominant over the viscous diffusivity ($\chi \gg \eta \gg \nu$). Unstable growth rates are 
shown as a function of the normalized vertical wavenumber $k_z V_{Az}/\Omega$. 
Model parameters are the same as those in Figure~\ref{fig3}.    }
\label{fig4}
\end{center}
\end{figure}

As seen from Figure~\ref{fig2}, the growth rate of axisymmetric MRI 
decrease with increasing the poloidal wavelength, 
but that of nonaxisymmtric MRI remains to be constant even at 
$k_z \lesssim \Omega/V_{Az}$. Thus, in multi-diffusive systems, 
the growth of nonaxisymmetric MRI dominates over that of axisymmetric one 
when the condition, $ \max (\lambda_\nu,\lambda_\eta) \gg \lambda_{\rm{MRI}}$ is 
satisfied. In this case, fast growing modes are limited to the "\textit{window A}" 
in Figures~\ref{fig2}a and \ref{fig2}b. In the case satisfying the condition, 
$\max (\lambda_\nu,\lambda_\eta) < \lambda_{\rm{MRI}}$, on the other hand, 
fast growing modes of axisymmetric and nonaxisymmetric MRI 
are limited to "\textit{window B}" in Figures~\ref{fig2}a and \ref{fig2}b, 
and their growth rates are comparable. 
Therefore, it is found that characters of fast growing modes of 
axisymmetric and nonaxisymmetric MRI are determined by the strength 
of vertical magnetic fields when the sizes of diffusive coefficients are fixed. 

Note that these properties appear under the situation that typical wavelength of 
nonaxisymmetric MRI is much longer than that of axisymmetric one. 
The toroidal magnetic fields are generally considerd to be 
dominant over the poloidal one in the stellar context (Heger et al.2005). 
Therefore, these properties would be important 
in the envelope of PNSs and solar radiative core. 

\section{APPLICATION TO PROTO-NEUTRON STARS}
In paper I, we have investigated the stability for the nonaxisymmetric MRI 
ignoring any diffusion processes and revealed that almost all regions of PNSs are unstable. 
However, we must consider the diffusive effects caused by the neutrino radiation 
in PNSs. The ohmic dissipation of hot nuclear matters 
can be also efficient in PNSs. Therefore, in this paper, the dispersion equation 
including multiple diffusion is derived. 
We apply it to the interiors of PNSs and reexamine the stability for the MRI. 
In the following, we restrict our discussion to the stably stratified regions of PNSs, 
which locates below the neutrinosphere 
(Janka \& M\"uller 1996; Thompson \& Murray 2000; Buras et al. 2003).  

\subsection{Physical Quantities in Proto-Neutron Stars}
\subsubsection{Dynamical Quantities}
During the core collapse, a significant amount of differential rotation 
can be generated in PNSs, even if the precollapse core rotate rigidly 
(Heger et al.2000,2005). Numerical studies of the rotating core-collapse 
suggest that the envelope of PNSs would rotate 
at the angular velocity $\Omega\simeq 100$--$1000 \rm{sec^{-1}}$ 
with shear rate $q \lesssim 1$ 
(Buras et al.2003; Kotake et al.2003; Villain et al.2004). 
These values strongly depend on the precollapse core models. 
Observations of young isolated pulsars associated with 
supernova remnants indicate that they rotate with $ \sim 100 \rm{sec^{-1}}$. 
Thus, we addopt $\Omega = 100 \rm{sec^{-1}}$ and $q = 1$ 
as typical rotational parameters. 

From the stellar evolution calculations, we can estimate the strength of 
magnetic fields in PNSs. 
Heger et al. (2005) study the evolution of a magnetized massive star 
with rotation, and show that the precollapse iron core has spatially 
homogenious megnetic fields. Toroidal magnetic components are amplified from 
poloidal one via various dynamo processes in the stellar evolutionary phases 
(Spruit 1999, 2002). The toroidal component is about 
$\sim 10^9$ G, which dominates over the poloidal one $\sim 10^6$ G.  
Assuming the conservation of magnetic flux ($B\propto\rho^{2/3}$), the envelope of the 
PNSs have $ B_\phi = 10^{13} $ G and $ B_z = 10^9 $ G. 
This would be the lower limit of the field strengths, because 
the magnetic fields could be amplified by turbulent dynamos or 
field wrapping during the core-collapse (Sawai et al.2005). 
Typical radius and density of PNSs are $R \sim 3\times 10^6\ \rm{cm}$ and 
$\rho \sim 10^{12} \ \rm{g\ cm^{-3}}$. 
The corresponding Alfv\'en frequencies $\omega_A$ and $\omega_{Az}$ are given by 
\begin{eqnarray}
&& \omega_A = 1.0\ \Big(\displaystyle{\frac{B_\phi}{10^{13} \ \rm{G}}}\Big) 
\Big( \displaystyle{\frac{R}{3\times10^6 \ \rm{cm}}} \Big)^{-1} 
\Big( \displaystyle{\frac{\rho}{10^{12}\ \rm{g\ cm^{-3} }}} \Big)^{-1/2} \ \ \rm{sec}^{-1} \label{eq49} \;, \\
&& \omega_{Az} = 10^{-4} \Big(\displaystyle{\frac{B_z}{10^{9} \ \rm{G}}}\Big) 
\Big(\displaystyle{\frac{R}{3\times10^6 \ \rm{cm}}}\Big)^{-1} 
\Big(\displaystyle{\frac{\rho}{10^{12}\ \rm{g\ cm^{-3}} }}\Big)^{-1/2} \ \ \rm{sec}^{-1} \label{eq50} \;. 
\end{eqnarray}

The buoyancy frequencies in the stably stratified layer of PNSs are 
sensitive to the microscopic physics such as the equation of state and leptonic fraction 
(Buras et al.2003; Thompson et al.2005; Dessart et al.2005). In general, 
the sizes of the thermal and leptonic buoyancy frequencies are 
much larger than the angular velocity ($N_T^2 \sim \ N_L^2 \gg \Omega^2$) and the 
both are of the same order below neutrinosphere (Janka \& M\"uller 1996). 
Thus, we assume that to be $N_T \sim  N_L = 10\Omega$ in this section. 

\subsubsection{Quadruple Diffusivities}
We estimate the various diffusion coefficients in PNSs. 
Under the equiriblium diffusion approximation, 
we can derive the heat and chemical diffusivities analytically as shown in \S~2. 
As the opacity $\kappa_0$, we adopt the simplified form derived by Janka (2001); 
\begin{eqnarray}
\kappa_0 &\sim& \displaystyle{\frac{\rho \sigma_0}{m_u}}
\displaystyle{\frac{\langle\epsilon_{\nu}^2\rangle}{(m_ec^2)^2}} f(\alpha_c,Y_n,Y_p)\nonumber \\
&=& 10^{-6} \Big( \displaystyle{\frac{\rho}{10^{12} \ \rm{g\ cm^{-3}}}} \Big) 
\Big(\displaystyle{\frac{k_B T}{4\ \rm{MeV}}}\Big)^2 \ \ [\rm{cm}^{-1}] \;. \label{eq53}
\end{eqnarray}
Here $\sigma_0 = 1.76\times10^{-44} \ \rm{cm}^2 $ is 
the characteristic weak interaction cross section, 
$m_e c^2 = 0.511\ \rm{MeV}$ is the rest-mass energy of the electron, $m_u$ is the atomic mass unit, 
$f(\alpha_c,Y_n,Y_p) \approx 1$ is a function of the vector coupling constant $\alpha_c$, 
and the number fractions of free neutrons, $Y_n$, and protons, $Y_p$. 
The contributions from the neutral-current scattering and the charged-current absorption 
of the electron neutrino and the anti-electron neutrino are included as the opacity source, 
but neutrino annihilation is not since the cross section is vanishingly small inside PNSs. 
Considering such the opacity, we can estimate the heat and chemical diffusivities as follows, 
\begin{eqnarray}
\chi &= & 10^{12} \Big(\displaystyle{\frac{\rho}{10^{12}\ \rm{g\ cm^{-3}}}}\Big)^{-2}\Big(\displaystyle{\frac{E_0}{10 \ \rm{MeV}}}\Big)^2
\Big(\displaystyle{\frac{k_BT}{4\ \rm{MeV}}}\Big)^{-1}\ \ [\rm{cm^2 sec^{-1}}] \;, \label{eq54} \\
\xi &= & 3\times 10^{11} \Big(\displaystyle{\frac{\rho}{10^{12}\ \rm{g\ cm^{-3}}}}\Big)^{-2}\Big(\displaystyle{\frac{E_0}{10 \ \rm{MeV}}}\Big)^2
\Big(\displaystyle{\frac{k_BT}{4\ \rm{MeV}}}\Big)^{-2}
\Big(\displaystyle{\frac{\mu_{\nu_e}}{4\ \rm{MeV}}}\Big)\ \ [\rm{cm^2 sec^{-1}}] \;. \label{eq55} 
\end{eqnarray}

The viscous stress is induced by the neutrinos in a hot nuclear matter, 
and it acts on the larger scale compared to the neutrino mean free path. 
When we assume the nondegenerate electron neutrino and anti-electron neutrino 
in degenerate nuclear matter, the heat diffusivity $\chi$ is related to 
the neutrino viscosity as 
\begin{eqnarray}
\nu &\sim & \displaystyle{\frac{c_p T}{\rho}} \chi 
= m_n^{2/3}\rho^{-2/3} T^2 [f(Y_p)]^{-1} \chi  \nonumber \\
& = &  10^{10} \Big( \displaystyle{\frac{\rho}{10^{12}\ \rm{g\ cm^{-3}}}} \Big)^{-8/3}
\Big(\displaystyle{\frac{E_0}{10\ \rm{MeV}}}\Big)^2
\Big(\displaystyle{\frac{k_BT}{4\ \rm{MeV}}}\Big) \ \ [\rm{cm^2 sec^{-1}}]\label{eq56} \;
\end{eqnarray}
(van den Horn \& van Weert 1984; Thompson \& Duncan 1993), 
where $c_p \approx \rho^{1/3}m_n^{2/3}T[f(Y_p)]^{-1}$ is the specific heat at constant pressure 
per volume, $f(Y_p) \sim 1$ is a specific function of the protons number fraction 
(Thompson \& Duncan 1993). This is because both of the momentum and heat are 
transported by neutrinos dominantly. 
 
The electric charge is transported via degenerate relativistic electrons there.
Momentum transfer rate is mainly determined by electron-proton scattering 
in the case not exceeding the standard nuclear density $\sim 10^{14} \ \rm{g\ cm^{-3}}$. 
When we consider the electron-proton collision, the electrical resistivity of hot nuclear matter 
is relatively low, $\mathcal{R} = 6\times 10^{-45} T^2\ \rm{sec}$ (Yakovlev \& Shalybkov 1991). 
Therefore, the magnetic diffusivity of the hot nucler matters is given by 
\begin{equation}
\eta = \displaystyle{\frac{c^2\mathcal{R}}{4\pi}}  = 10^{-3} \Big( \displaystyle{\frac{k_B T}{4 \ \rm{MeV}}} \Big)^2\ \ [\rm{cm^2 sec^{-1}}]\;, \label{eq57} 
\end{equation}
The magnetic diffusivity is extremely smaller than the other diffusion coefficients in PNSs. 

\subsection{Stability of Proto-Neutron Stars}
We discuss the characteristics of the axisymmetric and 
nonaxisymmetric MRI in the stably stratified regions of PNSs. 
In what follows, we use the physical quantities derived 
in \S~4.1; $q=1$, $\Omega=100\ \rm{sec}^{-1}$, 
$\omega_A = 1.0\ \rm{sec}^{-1}$, $\omega_{Az}=10^{-4}\ \rm{sec}^{-1}$, and 
$N_T = N_L = 10^3\ \rm{sec^{-1}}$. 
As the diffusivities, we addopt the values of 
$\chi=10^{12}\ \rm{cm^2 sec^{-1}}$, $\xi=3\times10^{11}\ \rm{cm^2 sec^{-1}}$, 
$\nu = 10^{10}\ \rm{cm^2 sec^{-1}}$, and $\eta = 10^{-3}\ \rm{cm^2 sec^{-1}}$. 
The features of MRI depend on the stellar merdian altitudes. 
Thus, we define the polar angle $\theta \equiv \tan^{-1} (r/z)$ 
which denote the direction of the entropy and leptonic gradients. 
Using a polar angle $\theta$, the radial and vertical components of 
the buoyancy frequencies are written as 
$N_{r} = N\sin\theta $ and $ N_{z} = N\cos\theta $.

\subsubsection{Stability of the equatorial region}
First of all, we discuss the growth of MRI at the equatorial plane of PNSs 
($\theta =\pi/2$) in order to show the diffusive effects clearly. 
As described in Paper I, the equatorial region is stable to the MRI 
in the diffusionless case. 
This is because the stabilizing effect of the negative buoyancy 
becomes the maximum at the equatorial plane ($N_{Te},N_{Le} \gg \Omega$). 
Considering the diffusion processes, the stabilizing effect is 
reduced and the growth of MRI could be revived even at the equatorial regions. 

\begin{figure}
\begin{center}
\begin{tabular}{c}
\scalebox{0.65}{\rotatebox{0}{\includegraphics{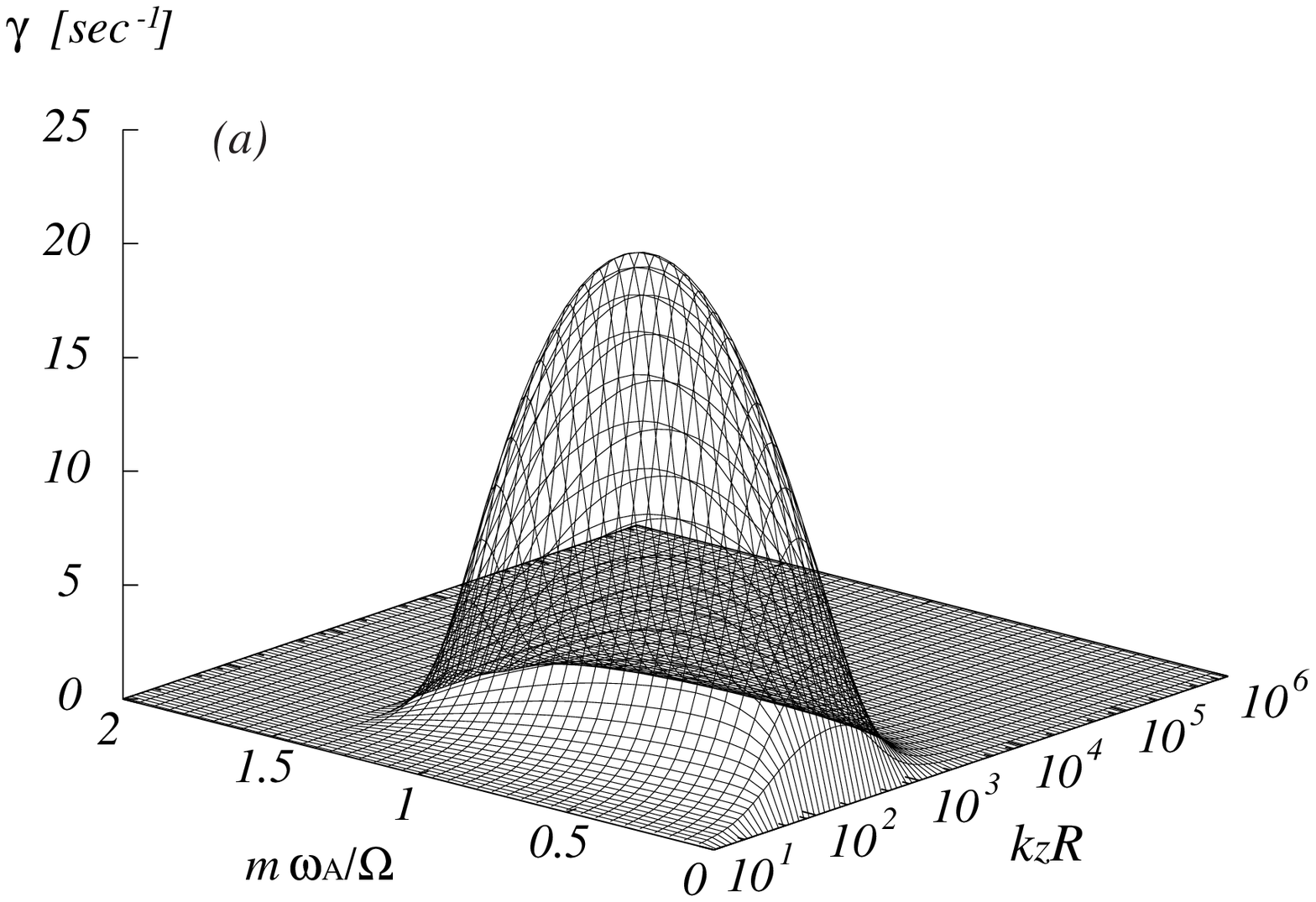}}} \\
\end{tabular}
\begin{tabular}{c}
\scalebox{0.6}{\rotatebox{0}{\includegraphics{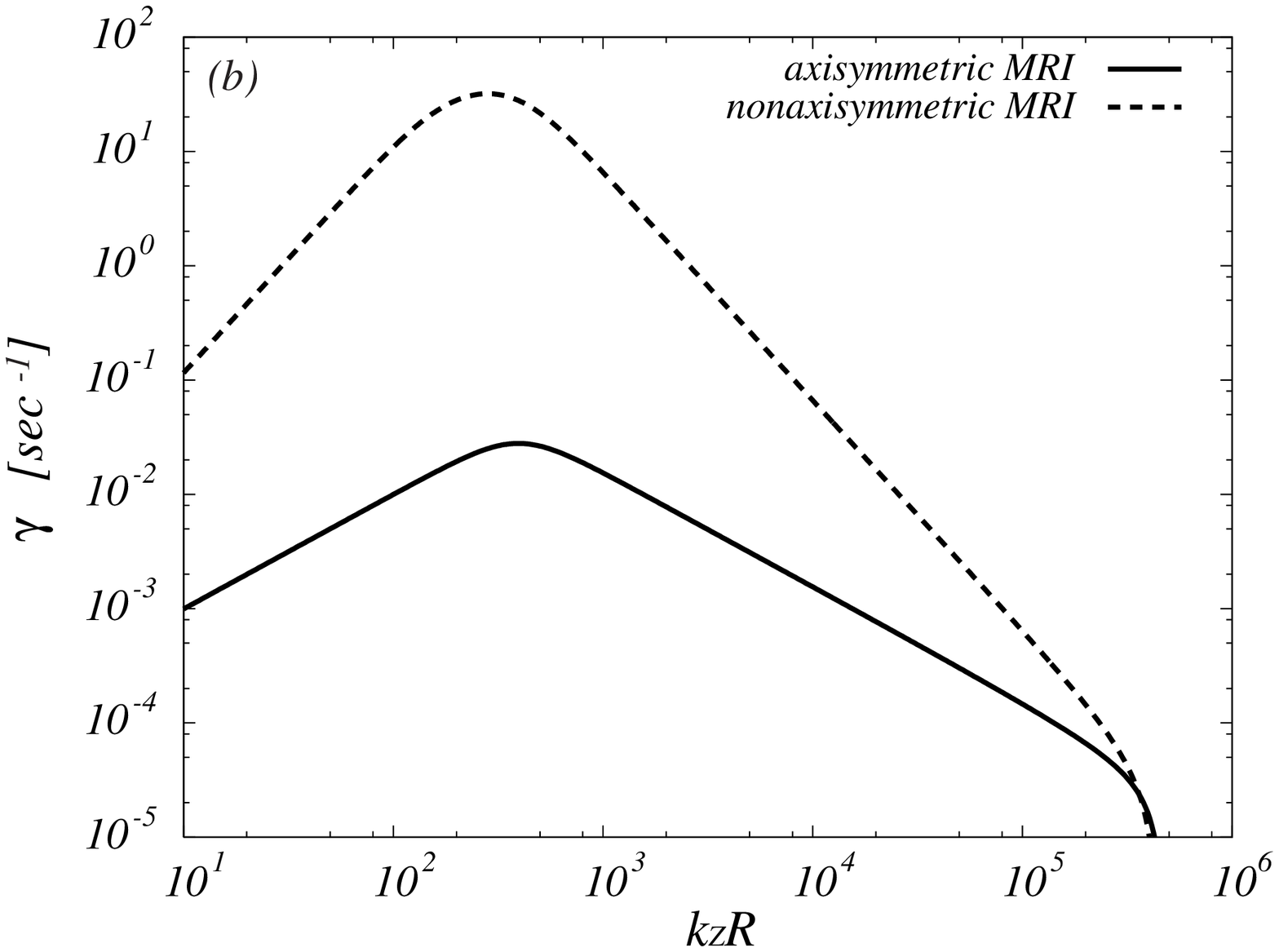}}} \\
\end{tabular}
\caption{(a) Three-dimensional plot of the unstable growth rate as a function of the 
azimuthal wavenumber $m$ and the normalized vertical wavenumber $k_z R$, and 
(b) slices of the three-dimensional plot at $m=0$ and $m=m_{\rm{max}}$. 
The size of the system is assumed to be $R = 3\times 10^6\ \rm{cm}$. 
The modes with $m=0$ are corresponding to the axisymmetric MRI (solid line), 
and the other is the nonaxisymmetric MRI (dashed-line). We choose the model parameters 
as $q=1.0$, $\Omega=100\ \rm{sec^{-1}}$, $N_T=N_L=10\Omega\ \rm{sec^{-1}}$, 
$\omega_A=1.0\ \rm{sec^{-1}}$, and $\omega_{Az}=10^{-4}\ \rm{sec^{-1}}$. 
Quadruple diffusivities are assumed as $\chi = 10^{12}\ \rm{cm^2\ sec^{-1}}$, 
$\xi = 3\times 10^{11}\ \rm{cm^2\ sec^{-1}}$, $\nu = 10^{10}\ \rm{cm^2\ sec^{-1}}$, and 
$\eta = 10^{-3}\ \rm{cm^2\ sec^{-1}}$, respectively. These figures represent 
the characteristics of MRI at the equatorial plane of the PNSs. }
\label{fig5}
\end{center}
\end{figure}

Figure~\ref{fig5}a shows the maximum growth rate of MRI ($k_r =0$) 
at the equatorial plane of PNSs as a fuction of the azimuthal and vertical wavenumbers. 
The vertical wavenumber is normalized by the size of 
the system $R$ ($= 3\times 10^6 \rm{cm}$). 
In the Figure~\ref{fig5}b, the slices of Figure~\ref{fig5}a 
at $m=0$ and $m=m_{\rm{max}}$ are depicted. 
From these figures, we find that the maximum growth rate of 
nonaxisymmetric MRI ($\sim 10\ \rm{sec^{-1}}$) is significantly larger than 
that of axisymmetric one ($\sim 0.01\ \rm{sec^{-1}}$). 

The range of poloidal wavelength which involve the fast growing modes of MRI 
is determined by the condition given by equation (\ref{eq47}). 
Substituting the physical quantities, we obtatin the upper and lower bounds of 
the wavelength for the fast growing modes in PNSs as follows;
\begin{equation}
10^4 \lesssim \lambda \lesssim 10^5 \ \ [\rm{cm}] \;. \label{eq58}
\end{equation}
This is surely consistent with the Figure~\ref{fig5}. 
The lower bound of this range is determined by the wavelength for 
the neutrino viscosity being effective. 
This means that the critical wavelength of axisymmetric MRI $\lambda_{\rm{MRI}}$ is 
smaller than $\lambda_\nu$. In this case, fast growing modes of MRI are limited to the 
"\textit{window A}" in Figure~\ref{fig2}. Therefore the growth of nonaxisymmetric MRI 
dominates over that of axisymmetric one (see \S~3.2.3). 

\subsubsection{Dependence of the maximum growth rate of MRI on stellar merdian altitudes}
\begin{figure}
\begin{center}
\begin{tabular}{c}
\scalebox{0.6}{\rotatebox{0}{\includegraphics{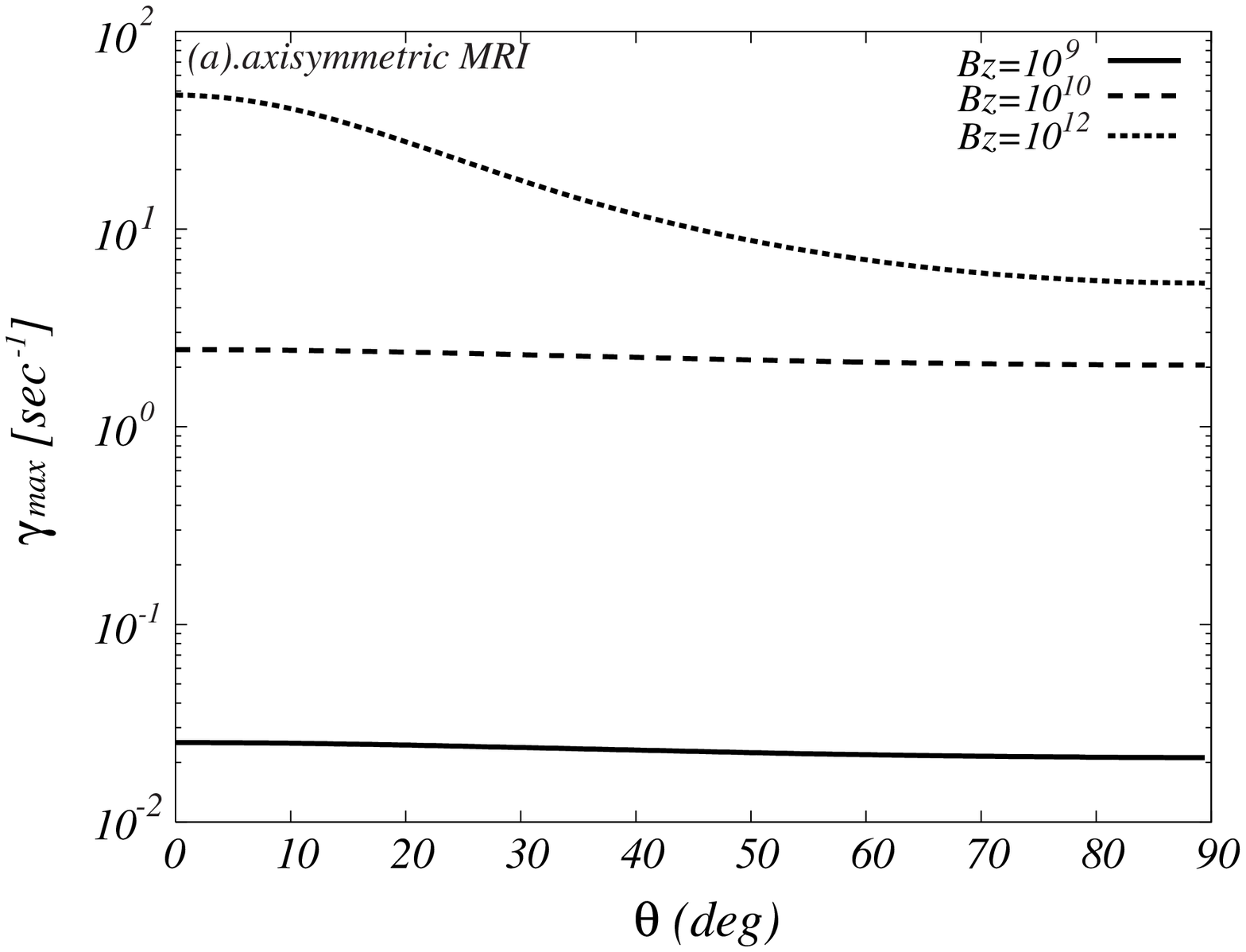}}} \\
\end{tabular}
\begin{tabular}{c}
\scalebox{0.6}{\rotatebox{0}{\includegraphics{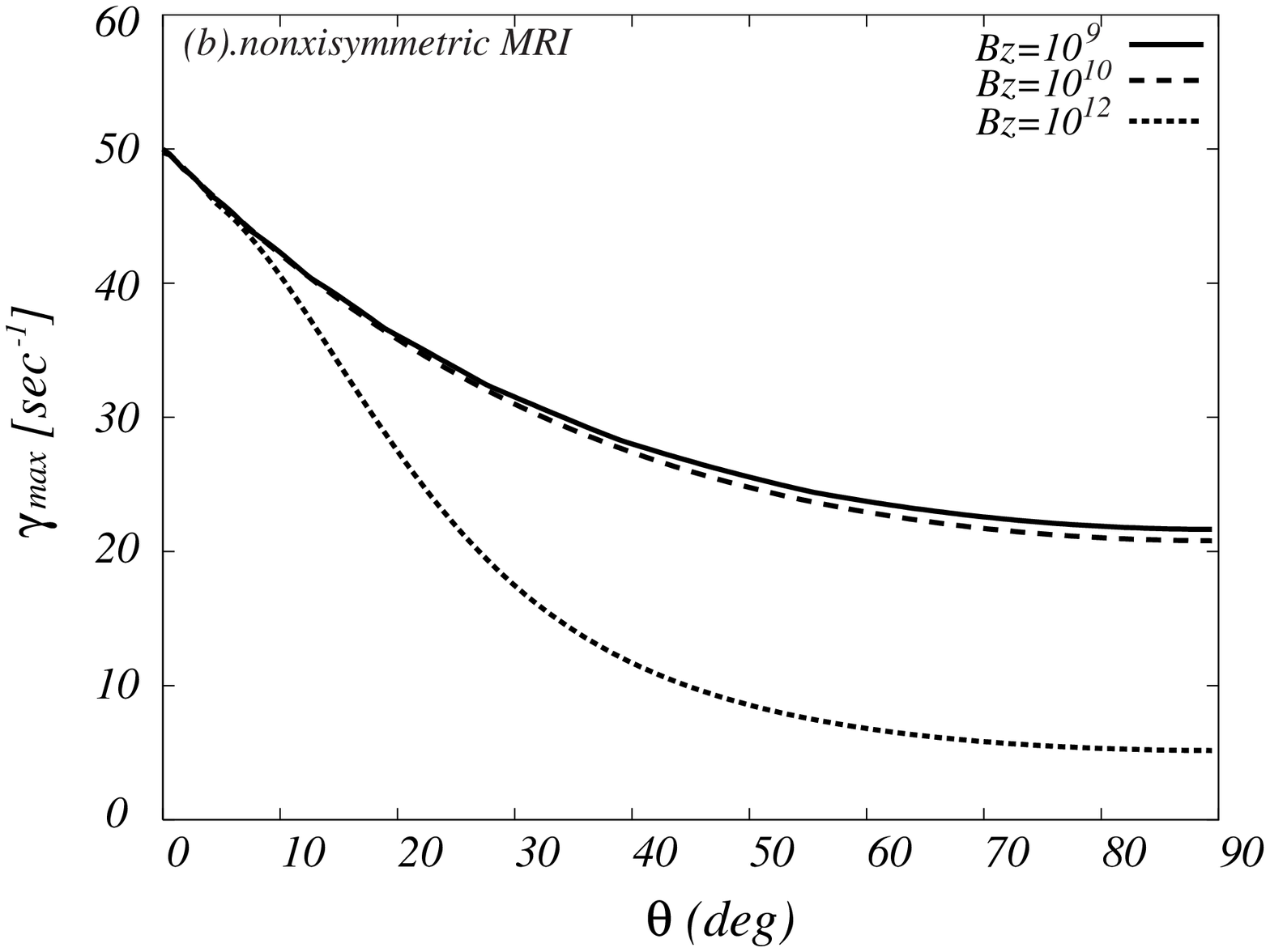}}} \\
\end{tabular}
\caption{The maximum growth rate of the MRI as a function of 
the polar angle $\theta$ for the cases of $B_z = 10^{9},10^{10}$, and $10^{12}$G. 
The upper panel shows the behavior of axisymmetric MRI ($m=0$), 
and the lower panel is nonaxisymmetric MRI ($m\ne0$). 
We search the fastest growing modes solving our simplyfied dispersion equation~(\ref{eq31}). 
The strength of toroidal fields is assumed to be constant in all cases, 
$B_{\phi}=10^{13}\ \rm{G}$. The other parameters are the same as those in Figure~\ref{fig5}. 
Note that the vertical axis of upper figure is measured by the logarithmic scale, 
but that of bottom figure is measured by the normal scale.}
\label{fig6}
\end{center}
\end{figure}

The dependence of the maximum growth rate of MRI on 
stellar meridian altitudes $\theta$ is described in Figure~\ref{fig6}.  
Upper and lower panels show the growth rate of axisymmetric and 
nonaxisymmetric MRI for the case with different field strengths 
$B_z = 10^9, 10^{10}$, and $10^{12}$ G. 
As seen from these figures, the maximum growth rate decrease with increasing the 
polar angle $\theta$. This is because the stabilizing effect due to 
the stable stratification increases with polar angle $\theta$ (see Paper I). 
In addition, we can find that the maximum growth rate of axisymmetric MRI is 
strongly dependent on the strength of poloidal magnetic fields. 
When the strength of poloidal magnetic fields is weak ($\sim 10^{9-10}$G), 
nonaxisymmetric modes of MRI grow dominantly at any $\theta$. 
As the strength of poloidal fields increases, 
the maximum growth rate of axisymmetric MRI increases, and then 
becomes comparable to that of nonaxisymmetric MRI at $\sim 10^{12}$G. 

For the case with weak poloidal fields ($B_z \sim 10^{9-10}$G), 
the typical wavelength of axisymmetric MRI $\lambda_{\rm{MRI}}$ is 
much smaller than $\lambda_\nu$. Thus the most growing mode of 
axisymmetric MRI is not included in the range defined by equation (\ref{eq58}). 
This corresponds to the case that the fast growing modes are 
limited to the "\textit{window A}" in Figure~\ref{fig2}. 
Therefore, the nonaxisymmetric modes of MRI grow dominantly in this case. 
As the strength of poloidal fields increases, 
the length of $\lambda_{\rm{MRI}}$ approachs to $\lambda_\nu$. 
This corresponds to the picture that the possible range of the fast growing modes 
shifts from "\textit{window A}" to "\textit{window B}" in Figure~\ref{fig2}. 
Thus the maximum growth rate of axisymmetric MRI increases when the strength of 
poloidal fields increases. 

For the fast growth of axisymmetric MRI as much as that of nonaxisymmetric one, 
it is necessary to satisfy the condition $\lambda_{\nu} \lesssim \lambda_{\rm{MRI}}$. 
Assuming the critical strength of poloidal fields satisfying this condition as $B_c$, 
we can derive the lower limit of poloidal fields for the fast growth of axisymmetric MRI;
\begin{equation}
B \gtrsim B_c \sim (\rho \Omega\nu)^{1/2} \simeq 10^{12} 
\Big(\displaystyle{\frac{\rho}{10^{12}\ \rm{g\ cm^{-3}}}}\Big)^{1/8}
\Big(\displaystyle{\frac{\Omega}{100 \ \rm{sec^{-1}}}}\Big)^{1/2} 
\Big(\displaystyle{\frac{k_B T}{4\rm{MeV}}}\Big) \ [\rm{G}]\;. \label{eq60}
\end{equation}
Here we use the specific form of neutrino viscosity defined by equation~(\ref{eq56}). 
The growth of axisymmetric MRI is drastically suppressed by 
the neutrino viscosity when the strength of the poloidal magnetic fields is 
weaker than $\sim 10^{12}$G. This critical value corresponds to 
that derived in Figure~\ref{fig5}a numerically. 

Note that, in the situation with realistic quadruple diffusions, 
the modes with zero radial wavenumber ($k_r =0$) are the fastest growing modes. 
This is a remarkable difference from the diffusionless case that studied 
in Paper I, in which the modes with $k_r/k_z = N_r/N_z$ are most unstable one. 
That is, unstable fluid elements can move against the stratification 
in neutrino loaded PNSs. This character suggests that the MRI itself could operate 
as the process leading to overturnig fluid motions (Kato 1992; 
Balbus \& Hawley 1994). 

As a result, we conclude that stably stratified envelope of 
PNSs are always unstable to the MRI. Especially, the maximum growth rate 
of nonaxisymmetric MRI is significantly larger than that of 
axisymmetric MRI unless the poloidal magnetic components 
are extremely strong ($\gtrsim 10^{12}$ G). 
Typical growth time of MRI is about $100\ \rm{msec}$. 
This is much shorter than the neutrino cooling time 
of PNSs ($\sim 10\ \rm{sec}$), and thus, it is expected that 
the nonlinear growth of MRI affect the neutrino luminosity 
via effective heating processes caused by 
overturnig fluid motions and/or the other magnetic processes. 

\section{DISCUSSION}
The nonlinear evolution of the MRI could amplify the magnetic fields and 
drive MHD turbulence in PNSs. On the other hand, the growth of other 
instabilities is also expected 
(Bruenn \& Dinneva 1996; Miralles et al. 2000, 2002; Bruenn et al. 2004). 
In particular, the radiative-driven magnetoacoustic instability, 
which is refered as "neutrino bubble", is focused as the seed process of 
the turbulent mixing and the mechanism for generating pulsar kicks 
(Blaes \& Socrates 2003; Socrates et al. 2005). 
We compare the growth rates of the MRI and the neutrino bubble instability, and 
derive the condition for the MRI dominating over the neutrino bubble instability. 

The growth rate of neutrino bubble instability is given by  
\begin{equation}
\gamma_{\rm{max}} \simeq 3\times 10^2\Big(\displaystyle{\frac{B_p}{10^{15}\ \rm{G}}}\Big)
\Big(\displaystyle{\frac{g}{10^{13} \rm{cm\ sec^{-2}}}}\Big)
\Big(\displaystyle{\frac{T}{4 \rm{MeV}}}\Big)^{-1}
\Big(\displaystyle{\frac{\rho}{10^{12} \rm{g\ cm^{-3}}}}\Big)^{-1/2} \ \ \rm{sec^{-1}}\;, \label{eq61}
\end{equation}
where $B_p$ is poloidal magnetic fields (Socrates et al. 2005).
Considering the typical growth rate of MRI ($\sim q\Omega$), 
we can derive the condition for the MRI dominating the neutrino bubble instability at 
the certain situation of $T\sim 4\ \rm{MeV}$, $\rho \sim 10^{12} \ \rm{g\ cm^{-3}}$, and 
$g \sim 10^{13}\ \rm{cm\ sec^{-2}}$: 
\begin{equation}
\Omega \gtrsim 3\times 10^{-7} q^{-1}B_p \;. \label{eq62}
\end{equation}
This condition suggests that the MRI dominates over the neutrino bubble instability 
in the weakly magnetized systems with a rapid differential rotaion. 

From the condition (\ref{eq62}), we can speculate the evolutionary scenario of PNSs. 
As PNSs should rotate differentially and be magnetized weakly right after core bounce, 
the growth of MRI exceeds that of the neutrino bubble instability. 
In the early evolutionaly phase, the nonlinear evolution of MRI initiates and sustains 
MHD turbulence. Thus, turbulent motions induced by MRI would enhance 
the neutrino luminosity sufficiently to yield explosions. 
MHD turbulence leads to angular momentum transport. 
Thus, the rotational configurations within PNSs evolve toward a rigid rotation. 
If the PNSs rotate rigidly, the neutrinosphere becomes oblate, 
and the neutrino luminosity could be enhanced in the polar region 
(Kotake et al.2003,2004). Thus, the polar region would be heated 
by neutrinos preferentially. 

After magnetic fields are amplified by the MRI and the rotation profile of PNSs 
becomes the rigid rotation, the growth rate of the neutrino bubble instability 
can be comparable to that of the MRI. At this evolutionary phase, 
the neutrino bubble instability can grow nonlinearly and drive turbulent mixing 
as is suggested in Socrates et al.(2005). Local luminosity enhancements, 
which preferentially occur in the regions of strong magnetic field, 
lead to a net global asymmetry in the neutrino emission and the young neutron star is propelled 
in the direction opposite to these regions. 
In addtion, the magnetic buoyancy instability may also grow in PNSs (Masada et al. 2006 in prep). 
To reveal the origin of neutron star kicks completely, we should study the nonlinear 
coupling effects of these instabilities in detail. 

\section{SUMMARY}
In this paper, we investigate the stabiliy of differentially rotating, 
magnetized PNSs, including the effects of multiple-diffusions. 
We derive the local dispersion equation 
for both axisymmetric and nonaxisymmetric perturbations 
and apply it to the interior of PNSs. Our findings are summarized as follows. 

1. The heat and leptonic diffusions reduce the restoring force due to 
stable stratification and increase the maximum growth rate. 
The viscous damping suppresses the growth of short wavelength 
perturbations and decrease the maximum growth rate. 
The ohmic dissipation stabilizes short wavelength perturbations. 
In the multi-diffusive systems, the fast growing modes of MRI can be seen 
only when the poloidal wavelength is 
$$
\max{( \lambda_\nu , \ \lambda_\eta , \ \lambda_{\rm{MRI}})} 
\lesssim \lambda \lesssim \min{ ( \lambda_\chi, \ \lambda_\xi ) }\;.
$$ 
The upper bound is determined by the lengthscale that the restoring force 
due to the stable stratification is reduced by the heat or leptonic diffusion. 
The lower bound is determined by the lengthscale for ohmic dissipation or 
viscous damping being effective. 

2. The necessary conditions of the stability for MRI in 
multi-diffusive case are derived from our simplified dispersion equation. 
The conditions are similar to that of Menou et al.(2004), 
in which only the stability for axisymmetric MRI is investigated.

3. MRI can grow in realistic neutrino loaded PNSs. 
For the case with $B_\phi \gg B_z$, 
the growth of nonaxisymmetric MRI dominates over axisymmetric one 
unless the vertical field is extremely strong ($B_z \gtrsim 10^{12} \rm{G}$).  
Typical growth time of fast growing MRI modes is about $100 \ \rm{msec}$, which is 
much shorter than the neutrino cooling time of PNSs. 

4. The condition that the growth of the MRI dominates over 
the neutrino bubble instability is given by 
$$\Omega \gtrsim 3\times 10^{-7} q^{-1}B_p \;,$$ 
at $T\sim 4\rm{MeV}$, $\rho\sim 10^{12}\rm{g\ cm^{-3}}$ and $g\sim 10^{13}\rm{cm\ sec^{-2}}$. 
Right after the core bounce, the growth of the MRI is dominant over the neutrino bubble instability. 
After the strong magnetic field is generated by the MRI and 
the rotational configuration shift to a rigid shape, 
the growth rate of the neutrino bubble instability could become larger than that of the MRI. 

\acknowledgments
We thank V. Urpin and S. Nagataki for helpful discussions. 
TS is supported by the Grant-in-Aid (16740111, 17039005) from the
Ministry of Education, Culture, Sports, Science, and Technology of
Japan.



\clearpage

\clearpage

\clearpage

\clearpage

\clearpage

\clearpage

\clearpage
\end{document}